\documentclass[%
 reprint,
aps,
pra,
]{revtex4-2}

\usepackage{amsmath}
\usepackage{amssymb}
\usepackage{color,soul}
\usepackage{graphicx}
\usepackage{dcolumn}
\usepackage{bm}
\usepackage{physics}
\DeclareMathOperator{\spn}{span}
\usepackage{hyperref}

\begin{document}

\preprint{APS/123-QED}

\title{Preparation of cavity Fock state superpositions by reinforcement learning exploiting measurement back-action}

\author{Arthur Perret}
\author{Yves Bérubé-Lauzière}%
 \email{Yves.Berube-Lauziere@USherbrooke.ca}
\affiliation{%
 Institut quantique and Département de génie électrique et de génie informatique,\\
 Faculté de génie, Université de Sherbrooke, Sherbrooke, Québec, J1K 2R1, Canada
}%

\date{\today}

\begin{abstract}

Preparation of bosonic and general cavity quantum states usually relies on using open-loop control to reach a desired target state. In this work, a measurement-based feedback approach is used instead, exploiting the non-linearity of weak measurements alongside a coherent drive to prepare these states. The extension of previous work on Lyapunov-based control is shown to fail for this task. This prompts for a different approach, and reinforcement learning (RL) is resorted to here for this purpose. With such an approach, cavity eigenstate superpositions can be prepared with fidelities over 98$\%$ using only the measurements back-action as the non-linearity, while naturally incorporating detection of cavity photon jumps.
Two different RL frameworks are analyzed: an off-policy approach recently introduced called truncated quantile critic~(TQC) and the on-policy method commonly used in quantum control, namely proximal policy optimization~(PPO). It is shown that TQC performs better at reaching higher target state fidelity preparation. 
\end{abstract}

\maketitle


\section{\label{sec:intro}Introduction}

Quantum states in the Hilbert space of microwave cavities have been at the forefront of recent efforts towards a fault-tolerant quantum computer~\cite{Joshi_2021,MA_2021,CAI_2021}. Improvements in the coherence times of cavities, along with the well understood and highly biased noise hints at a scalable path using such devices. Indeed, a whole class of bosonic codes - that is, encoding information in the multiple energy levels of harmonic oscillators - is being developped~\cite{Micheal_2016,Grimsmo_2020}. 

Preparation of states with such encodings can however be challenging, using for instance dissipative or adiabatic approaches for the cat states~\cite{Grimm_2020,Zhou_2022,Mirrahimi_2014}. More general bosonic states,
such as binomial code states or arbitrary superpositions, rely on open-loop protocols~\cite{law_1996, hofheinz_2008, Hofheinz_2009, krastanov_2015, heeres_2015, heeres_2017, fosel_2020}. While arbitrary states can be prepared this way, these techniques do not exploit the advantages made possible by feedback approaches, namely their added robustness. Indeed, even if full unitary control is available, there is no guarantee that one can recover from an error such as a cavity jump occuring during state preparation. By repeatedly monitoring the cavity state during state preparation, one can adjust the control action as necessary in a feedback loop, without using more physical resources than for an open-loop approach.

Deterministic state preparation using feedback has been demonstrated through the pioneering work of Haroche's group, whereby cavity eigenstates were prepared using the back-action from weak measurements as a decimation procedure~\cite{dotsenko_2009, sayrin_2011}. Similar approaches have also been developed for both superconducting and semiconductor qubits, using either classically-inspired feedback techniques~\cite{aarab_2022} or reinforcement learning approaches~\cite{Porotti2022a, Reuer_2022}. In these works, however, only eigenstates are being prepared. Recently, Porotti \textit{et al.}~\cite{Porotti2022a} extended this idea using continuous-time multiplexed measurements~\cite{Essig_2021} to prepare both single Fock states and superpositions. Preparing superpositions, however, proved to be more challenging with relatively low state preparation fidelities, and was restricted to a limited set of states.

The present work focuses on preparing cavity state superpositions in a circuit quantum electrodynamic (cQED) architecture~\cite{Blais_2004, Blais_2021, Wallraff_2004}, using a generalized version of the measurements used by Haroche's group. It is shown that careful choice of these measurements, and therefore of their back-action on the cavity's state, allows reaching target states with high fidelity, but at the cost of a more elaborate optimization procedure. Indeed, Haroche's method crucially relied on having the cavity Fock states being the fixed points (eigenstates) of the measurement operators. The control problem was therefore simplified to bring the current cavity state near the targeted fixed point, with the measurement back-action playing the role of an attractor, hence allowing asymptotic convergence~\cite{Amini_2013}.
Extending Haroche's approach to prepare superpositions of cavity eigenstates is, however, not direct. As discussed in references~\cite{mirrahimi_stabilizing_2007, vanHandel_2005, liu_lyapunov-based_2017, Cong_2013}, degeneracies of the measurement operators eigenvalues, while necessary to stabilize superpositions, makes wholes subspaces invariant under these operators. In such cases, relying on the measurements back-action alone will only ensure convergence towards the subspace containing the target state. To counteract this, one would need to add a second Hamiltonian control term to lift the degeneracies~\cite{liu_lyapunov-based_2017}. However, additional control introduces new error pathways, and makes the whole control protocol more complex. Such an approach will thus not be pursued here.
Since previous approaches can only converge to subspaces, this raises the following question: Is it possible to find a controller that allows convergence to specific state superpositions with high fidelity without additional Hamiltonian control terms? This is the question addressed in the present work.

A further limitation in the present work is that only coherent displacements can be applied to drive the cavity. 
The limited control drive problem
is interesting in its own right, as it
puts forward the question of how to interact in an optimal way with the quantum back-action resulting from repeated weak measurements. Recent work focusing on feedback-based quantum control~\cite{Fosel_2018, Sivak_2022, Zhikang_2020,Borah_2021, Porotti2022a} has begun to address similar questions 
with deep reinforcement learning~(RL) through the underlying formalism of quantum-observable Markov decision processes~\cite{Barry_2014}, the quantum analogue of the classical RL framework. In the present work, the large state space along with the high variance of the dynamics coming from random measurement outcomes makes the preparation of state superpositions with RL challenging. Another objective of the present paper is thus also to pinpoint an RL method able to handle such dynamics and provide for convergence to specific desired cavity state superpositions.

The rest of this paper is structured as follows: In Sect.~\ref{sec:SystemModel}, the cQED model considered is defined with its allowed measurements and controls. We develop the theory showing that with the available measurements, only certain subspaces of superpositions can be stabilized. Considering this limitation, the control problem is then considered in the light of RL. Sect.~\ref{sec:results} presents results for the ideal decoherence free setting as well as when decoherence is present. While decoherence reduces the state preparation fidelities that can be achieved, state-of-the-art devices operate in a regime where our approach allows the stabilization and recovery from photon jumps in the cavity. In Sect.~\ref{sec:discussion}, the learned behavior of the RL agent is analyzed, comparing its optimal policy with other approaches, namely an on-policy RL agent along with a Lyapunov function-based controller.

\section{System definition and control approach}
\label{sec:SystemModel}

\subsection{System considered}
\label{sec:Model}

A standard cQED architecture is considered here, in which a qubit is dispersively coupled to one mode of a microwave cavity, with the following Hamiltonian
\begin{equation}
H_{system} = \omega_{c} N + \frac{\omega_{q}}{2}\sigma_{z} +   \chi N \:\sigma_{z} ,
\end{equation}
subject to a resonant control drive on the cavity given by
\begin{equation}
H_{control} = i(\varepsilon a^\dagger - \varepsilon^* a).
\end{equation}
Here, $N = a^{\dag}a$ is the number operator, with $a$ and $a^\dagger$ being respectively the annihilation and creation operators of the cavity mode, and $\chi$ is the dispersive coupling strength between the qubit and the cavity mode. Associated with the control Hamiltonian is the displacement operator given by
\begin{equation}
  D(\alpha) = e^{\alpha a^\dagger - \alpha^* a} ,
\label{eq:displacement_op}
\end{equation}
with $\alpha = \varepsilon t$, the coherent resonant drive amplitude. Assuming the cavity to be in state $|\psi\rangle$, then after application of the resonant control drive it becomes $|\psi'\rangle = D(\alpha) |\psi\rangle$. In the case the density operator $\rho$ is used to specify the state rather than the state vector $|\psi\rangle$, then the state $\rho'$ after application of the drive is given by
\begin{equation}
    \rho' = D(\alpha) \rho D(\alpha)^\dagger = D(\alpha) \rho D(-\alpha) .
\label{eq:DispCavState}
\end{equation}
This transformation of $\rho$ into $\rho'$ can be written in superoperator form as~\cite{dotsenko_2009}
\begin{equation}
  \rho' = \mathbf{D}(\alpha) \rho . 
\label{eq:CohDriveDensMatSuperopForm}
\end{equation}

In the dispersive regime, no energy exchange occurs between the qubit and the cavity mode. Instead, the cavity experiences a light shift depending on the state of the qubit, while the latter experiences a phase shift depending on the cavity's state. As detailed below, this interaction is at the core of the present proposal, with the phase information acquired by the qubit updating the knowledge about the cavity's state. It will be seen that the back-action from qubit measurements then acts as a decimation procedure to prepare a target quantum state in the cavity~\cite{sayrin_2011}. This back-action is thus a resource for controlling the cavity.

\begin{figure}[t]
\includegraphics[width = 8.5cm]{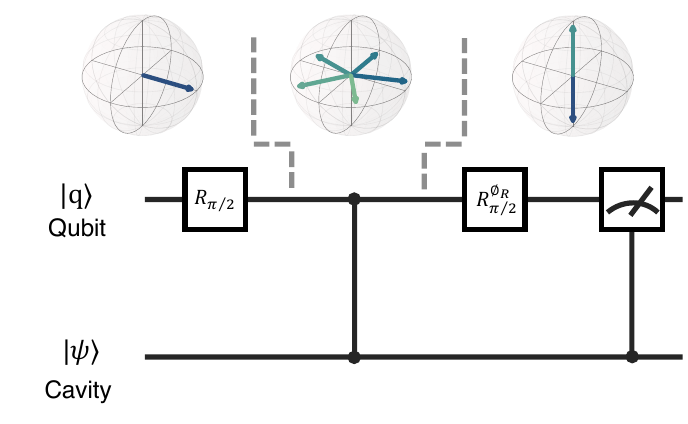}
\caption{Schematic of the measurement scheme for a single feedback cycle. A $\pi/2$ Ramsey pulse is applied to the qubit initially in state $|e\rangle$ to prepare the qubit's $\ket{+}$ state. After interaction between the qubit and cavity for a given time (interaction time), the qubit's state, which has now become dependent on the cavity's state that prevailed prior to the interaction, is projected back onto its energy eigenbasis, and subsequently measured. This Ramsey protocol implements a weak measurement of the cavity's state.}
\label{fig:meas_scheme}
\end{figure}

The measurement scheme's full sequence, as shown in Fig.~\ref{fig:meas_scheme}, starts by preparing the qubit in state $\ket{+} = \frac{\ket{e} + \ket{g}}{\sqrt{2}}$ by applying a $\pi/2$ pulse on it, where $\ket{g}$ and $\ket{e}$ are respectively its ground and excited states, and then letting the qubit interact with the cavity for a given interaction time $t_{int}$. After the interaction, the qubit and cavity evolve to the state
\begin{equation}
|\Psi \rangle
= \frac{1}{\sqrt{2}}
   \bigl(
   |g\rangle \otimes e^{-\frac{i}{2}(\phi_0 a^{\dag}a + \delta\phi)} \ket{\psi}
 + |e\rangle \otimes e^{\frac{i}{2}(\phi_0 a^{\dag}a + \delta\phi)} \ket{\psi}
   \bigr) ,
\end{equation}
where $\phi_0 = t_{int}\cdot \chi$ is the phase shift per photon present in the cavity, which is an experimentally tunable parameter, $\delta\phi$ is a constant phase shift whose exact form is not important here, and $\ket{\psi}$ is the state of the cavity. Applying a second $\pi/2$ pulse projects the qubit back onto its energy eigenbasis, with the measurement probabilities determined by the phase between the two superposed qubit eigenstates, a procedure known as Ramsey interferometry. In the case the cavity is in one of its eigenstate $\ket{n}$ (number state or Fock state), the state of the qubit after interaction can be written as
\begin{equation}
    |q\rangle = \frac{\ket{e} + e^{i(\phi_0 n + \delta\phi)}\ket{g}}
                     {\sqrt{2}} .
\label{eq:QubitStateSuperposCavState_n}
\end{equation}
This can be seen as the cavity state $\ket{n}$ imparting a phase $\phi_0 n + \delta\phi$ to the initial qubit superposition $\frac{(\ket{e} + \ket{g})}{\sqrt{2}}$. To each cavity eigenstate $\ket{n}$, there corresponds a direction on the equatorial plane of the qubit's Bloch sphere, this direction being specified by the angle $\phi_0 n + \delta\phi$.
In this case, Ramsey interferometry allows the weak QND measurement of the number of photons in the cavity~\cite{Haroche_2006}. 
Such weak measurements combined with coherent drive excitation of the cavity was the building block of the iterative measurement-based quantum feedback (MBQFB) control approach pioneered by Serge Haroche~\textit{et al.} to prepare cavity Fock states using atoms in Rydberg states as probe qubits~\cite{dotsenko_2009,sayrin_2011} (also called ancilla or auxiliary qubits in the error-correction literature). 

\subsection{Stabilization scheme}

Owing to the dispersive interaction between the microwave cavity and the qubit, the back-action on the cavity's state following a measurement of the state of the qubit giving as outcome $e$ or $g$ is obtained through the following measurement operators 
\begin{align}
   M_{g} &=  \cos\big( \frac{\phi_0 N - \varphi_R}{2} \big), \label{eq:meas_ops_g} \\
   M_{e} &= \sin\big( \frac{\phi_0 N - \varphi_R}{2} \big). \label{eq:meas_op_e}
\end{align}
Here, $\varphi_R = \phi_R - \delta\phi$, where $\phi_R$ is the phase of the second $\frac{\pi}{2}$ Ramsey pulse relative to the first one, which is a tunable parameter. More precisely, assuming the cavity to be in state $\ket{\psi}$ prior to a measurement of the qubit, and given that the qubit is measured to be in state $s$, with $s = e$~or~$g$, the state of the cavity after the qubit measurement is $\ket{\psi'} = M_s \ket{\psi}$. If the density operator is used instead of the state vector, this becomes
\begin{equation}
  \rho' = \frac{M_s \rho M_s^\dagger}{\tr(M_s \rho M_s^\dagger)}
\end{equation}
which in superoperator form is written as
\begin{equation}
  \rho' = \mathbf{M}_s \rho .
\end{equation}

To prepare a cavity Fock state superposition with an iterative feedback loop protocol in which a measurement operator $M_e$ or $M_g$ affects the cavity's state in each loop, similarly to the protocol of Haroche \textit{et al.}. for preparing single Fock states, it is necessary that the targeted Fock state superposition be left unchanged by the measurement's back-action so as to keep stable the state reached upon convergence of the feedback loops. This means that the Fock state superposition must be an eigenstate of both measurement operators.
For simplicity, taking as target the two-state superposition
\begin{equation}
  \ket{\psi^\mathrm{target}} = c_1 \ket{n_1} + c_2 \ket{n_2},
\end{equation}
where without loss of generality $n_2 > n_1$, the following conditions must therefore hold:
\begin{equation}
\begin{split}
    M_g \ket{\psi^\mathrm{target}} &= \lambda_g \ket{\psi^\mathrm{target}} , \\
    M_e \ket{\psi^\mathrm{target}} &= \lambda_e \ket{\psi^\mathrm{target}} .
\label{eq:CondEigVectMeasOps}
\end{split}
\end{equation}
These conditions translate to
\begin{equation}
\begin{split}
    \cos\big(\frac{\phi_0 n_1 - \varphi_R}{2}  \big) &= \cos\big(\frac{\phi_0 n_2 - \varphi_R}{2}  \big) , \\
    \sin\big(\frac{\phi_0 n_1 - \varphi_R}{2}  \big) &= \sin\big(\frac{\phi_0 n_2 - \varphi_R}{2}  \big) .
\end{split}
\end{equation}
Since both cosines and sines must be equal, the arguments must be equal to within $2\pi k $, with $k$ an integer, which imposes that the phase shift per photon be of the following form:
\begin{equation}
    \phi_0 = \frac{4 \pi k}{\Delta n},
\label{eq:modulo_condition}
\end{equation}
with $\Delta n = n_2 - n_1$ ($k = 1$ is used in the sequel).
It is to be noted that only $\Delta n$ matters in $\phi_0$. Furthermore, it is easily seen that the argument for the two-state superposition is readily generalizable to an arbitrary superposition of Fock states whose numbers differ by $\Delta n$. Thus, entire subspaces containing Fock states with numbers differing by $\Delta n$ can be stabilized. Specifically, for a given $\Delta n$ that can be chosen, leading to a specific value of $\phi_0$, states in the following subspaces can be stabilized by the measurement operators
\begin{equation}
\begin{split}
  W_0^{\Delta n}
    &= \spn \left\{ \ket{0}, \ket{\Delta n}, \ket{2\Delta n}, \ldots \right\}, \\
  W_1^{\Delta n}
    &= \spn \left\{ \ket{1}, \ket{1+\Delta n}, \ket{1+2\Delta n}, \ldots \right\}, \\
    &\vdots \\
  W_m^{\Delta n}
    &= \spn \left\{ \ket{m}, \ket{m+\Delta n}, \ldots \ket{m+l\Delta n}, \ldots \right\}, \\
    &\vdots \\
  W_{\Delta n - 1}^{\Delta n}
    &= \spn \left\{ \ket{\Delta n - 1}, \ket{2\Delta n - 1}, \ket{3\Delta n - 1}, \ldots \right\} .
 \end{split}
\end{equation}

These subspaces $W_m^{\Delta n}$, $m = 0, \ldots, \Delta n - 1$, will be called the \textit{stabilizable subspaces}.
It is seen that each basis state in the generic subspace $W_m^{\Delta n}$ contains a number $n$ of photons such that $n \!\! \mod \Delta n = m$.
Only superpositions of Fock states living inside each of these subspaces can be prepared using the iterative measurement feedback protocol described above. While the conditions given in Eq.~\eqref{eq:CondEigVectMeasOps} restrict the state superpositions that can be prepared, the modulo nature of the number of photons contained in the stabilizable subspaces is a resource that is exploited in many bosonic codes that allow for error correction\cite{Grimsmo_2020}.

\begin{figure}[h]
\includegraphics[width = 6cm]{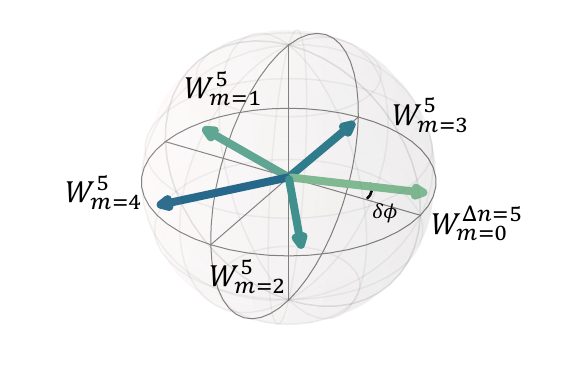}
\caption{Example of the mapping between stabilizable subspaces $W_m^{\Delta n}$ and directions in the equatorial plane of the qubit's Bloch sphere in the case $\Delta n = 5$ ($\phi_0 = 4\pi/5$). }
\label{fig:MapStabilizSubspBlochSph}
\end{figure}

Similarly to the discussion following Eq.~\eqref{eq:QubitStateSuperposCavState_n} about the effect on the qubit state superposition when the cavity is in a Fock state, here any cavity state in a stabilizable subspace $W_m^{\Delta n}$ imparts the same phase $\Phi^{\Delta n}(m)$ on a qubit superposition. Indeed, the qubit state after interaction with the cavity in such a state is given by
\begin{equation}
    |q_m^{\Delta n}\rangle
    = \frac{\ket{e} + e^{i\Phi(W_m^{\Delta n})}\ket{g}}
           {\sqrt{2}} ,
\end{equation}
with $\Phi\left( W_m^{\Delta n} \right) \equiv \Phi(m+l\Delta n) = \Phi^{\Delta n}(m) = \frac{4 \pi k m}{\Delta n} + \delta\phi \pmod{2\pi}$.
Hence, each subspace $W_m^{\Delta n}$ is mapped to its own direction in the equatorial plane of the qubit's Bloch sphere, since all basis vectors $\ket{m+l\Delta n}$ in this subspace are mapped to the same angle $\Phi^{\Delta n}(m)$, which does not depend on $l$.
Fig.~\ref{fig:MapStabilizSubspBlochSph} illustrates an example for $\Delta n = 5$.
It is to be noted that in the case $\Delta n = 2 p$ is even, with $p$ a positive integer, the subspaces with indices $m$ and $m + \Delta n / 2$ will be mapped to the same angle $\Phi^{\Delta n}(m)$ (mod~$2\pi$) in the equatorial plane. Furthermore, if $\Delta n = 2 p$, with $p$ even, that is $\Delta n = 4 r$, with $r$ a positive integer ($r = 1, 2, \ldots$), stabilizable subspaces with indices $m$ and $m + \Delta n / 4$ will be mapped to angles $\Phi^{\Delta n}(m)$ opposing in the equatorial plane (\textit{i.e.} angles that differ by $\pi$ (mod~$2\pi$)).
Since the case $\Delta n = 4 r$ is a particular case of $\Delta n = 2 p$, it means that in the case $\Delta n = 4 r$, four subspaces, namely the subspaces with indices $m$, $m + \Delta n / 4$, $m + 2 \Delta n / 4$ and $m + 3\Delta n / 4$ will be mapped on the same angle or opposing angle (in fact two on the same angle and the other two on the opposing angle).
The significance of this is that it is not possible to discriminate subspaces which are mapped to the same angle (mod~$2\pi$) through the probabilities of the measurement outcomes $e$ or $g$ when measuring the qubit.
However, as discussed below, by an appropriate choice of $\varphi_R$, subspaces that are mapped to opposing angles, can be discriminated through the probabilities of the measurement outcomes for the qubit (note that for example if $\varphi_R$ is taken equal to $\pi/2$, that is when the Ramsey interferometer operates at mid-fringe as is often done~\cite{dotsenko_2009}, then one cannot discriminate subspaces with opposing angles).

\subsubsection*{Odd and even stabilizable subspaces}

A $4\pi$ factor appears in the numerator of Eq.~\ref{eq:modulo_condition}, as opposed to the $2\pi$ more often seen in usual parity measurements~\cite{Sun2014}. This is caused by the necessity of having the same eigenvalue for all superposed states after a measurement, rather than preserving the same measurement probability for these states. Indeed, setting the prefactor as $2\pi$ would imply to keep track of an alternating phase between each measurement for the corresponding Fock states which have non-zero population. Such tracking of the phase is possible in superconducting circuits, as it is always possible to know when a system measurement is carried out.

To determine the value of the parameter $\phi_0$ in the measurement operators, this implies that setting a $4\pi$ factor for odd subspaces leads to an adequate  subspace stabilizer, with no overlapping problem. For even subspaces, the only option is to choose a $2\pi$ factor to prevent the overlapping, and then resort to phase tracking.

The second parameter defining the measurement operators, $\varphi_R$, is free. Dotsenko \textit{et al.}~\cite{dotsenko_2009} is followed here, and its value is set at mid fringe visibility for odd $\Delta n$ subspaces. On the Bloch sphere representation, this means the projection axis is perpendicular to the target subspace. For even subspaces, each vector on the Bloch sphere is always opposing another one, which prevents the use of a perpendicular projection axis. Rather, $\phi_R$ is set at an angle of $\frac{2\pi}{5}$ from the target subspace, a value chosen so that all subspaces can be assigned a different probability of being measured in either $\ket{g}$ or $\ket{e}$.

\subsection{Control decision problem}

As mentioned in the introduction, conventional control techniques to stabilize specific superpositions such as Lyapunov-based control are  difficult to implement, and are bound to get trapped in local minima, as the measurement back-action stabilizes the whole subspace consisting of superpositions of states with $n$~mod~$\Delta n$ photons. In this context, designing a Lyapunov function that decreases motonically towards the target state rather than the target subspace is not straightforward, if possible at all. Such an approach would prevent both large early displacements in the feedback sequence that temporarily move the state further from the target, as well as small displacement corrections within the target subspace.

To overcome this difficulty, a reinforcement learning (RL) approach is used herein, which in principle can learn the system dynamics directly in a model-free manner from the experimental apparatus.

While previous work on feedback RL-based quantum control in continuous actions spaces has mainly focused on on-policy architectures~\cite{Porotti2022a,Sivak_2022, Borah_2021}, here use is made of the more sample efficient off-policy learning paradigm. Specifically, an actor-critic architecture is resorted to, where the actor network updates its policy by deriving optimal state-action tuples from the Q-function learned by the critic. 
In contrast to Monte Carlo based methods such as proximal policy optimization~(PPO), the critic here bootstraps state-action estimates in the update of its own Q-function, which reduces the variance in the reward estimate. However, this also increases sensitivity to bias in the estimator, which can be problematic in stochastic setting, and lead to overestimation of the Q-function~\cite{Thrun_1993}. 
To prevent this, a distributional RL approach~\cite{bellemare_2017, bellemare_2022} is used here, which learns the Q-function by regressing over the distribution of the returns, rather than their mean. Although it is still unclear what makes such an approach more stable, it is believed to help Q-learning when using non-linear functions approximation~\cite{Lyle_2019}, by providing richer information about the environment dynamics. Additionally, a recent algorithm has been introduced, called truncated quantile critic~(TQC), which builds a distributional version of the Q-function on top of a soft-actor critic algorithm. TQC allows for finer control of overestimation bias, and has demonstrated superior performance compared to other state-of-the-art algorithms, particularly in high-dimensional stochastic environments~\cite{kuznetsov_2020}. Here, the TQC implementation from Stable-Baselines3~\cite{stable-baselines3} is resorted to, whereas the specific hyperparameters used can be found in Appendix~\ref{sec:appA}. 

\begin{figure}[h]
\includegraphics[width = 8.5cm]{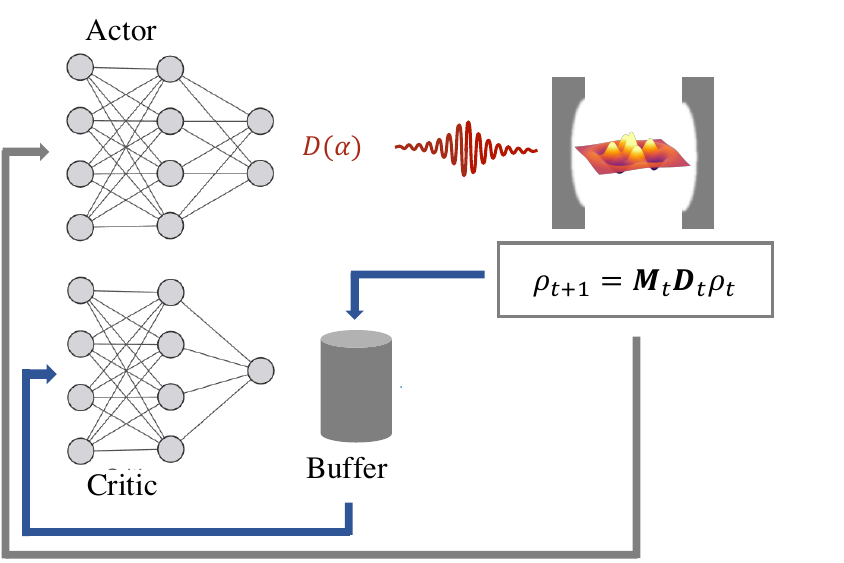}%
\caption{Schematic of the RL procedure. The cavity state is estimated by a quantum filter $F$, which is fed as input to the actor. The actor then outputs a displacement amplitude to be applied to the cavity. At every step, the critic samples from a buffer of past experiences to approximate the Q-function. During policy iteration, the actor optimizes its policy with the Q-function learned by the critic.}
\label{fig:control_framework}
\end{figure}

Finally, the reward function must be defined. Its formulation is critical in guiding the agent towards achieving maximum fidelity with respect to the target state, using as few feedback cycles as possible. 
Similar to Porotti \textit{et al.}~\cite{Porotti2022a}, it was found, for the present purposes, that a reward consisting of higher powers of the fidelity was most effective to prevent convergence to a local optimum. Defining the fidelity between two density matrices as
\begin{equation}
    F(t) = \left(\tr\sqrt{ \sqrt{\rho(t)} \rho^\mathrm{target} \sqrt{\rho(t)} }  \right)^{2},
\label{eq:FidelitySqrRoots}
\end{equation}
the reward function is chosen to be
\begin{equation}
    r(t) = \big(F(t)\big)^{4} + 4\; \big(F(t)\big)^{25} .
\label{eq:RewFct}
\end{equation}

The choice of the form of the reward function was guided by heuristics, and the specific numerical values of the exponents
(4 and 25)
and coefficients
(1 and 4)
were empirically chosen so as to give agents which reach highests fidelities. The right-hand side of Eq.~\eqref{eq:RewFct} includes two terms that influence the training process. The first term helps to accelerate training by providing a dense reward to the agent, whereas the second ensures that the maximum reward available is sharply peaked near unit fidelity. This borrows from approaches found in curriculum learning~\cite{Bengio_2009,Ma_2022}, where the agent is given guidance in a first step on how to reach the correct subspace, and then in a second step on how to reach the actual target state inside this subspace.

\subsection{Quantum filter}

To use a neural network to process the sequence of measurement outcomes
$e$ or $g$,
it is necessary to incorporate some form of memory of past inputs into the network. One option is to use a recurrent neural network, which can process sequential data by using feedback connections that allow information to be passed from one time step to the next.

Alternatively, the density matrix of the cavity state can be used as an input vector, which encodes all past information about the system. This approach requires using a quantum filter to estimate the cavity state recursively, using the displacement drive $\alpha$ and measurement outcome
as inputs at each time step. A quantum filter, which provides a state estimator analogous to a Kalman filter in classical control theory, allows obtaining in the computer an estimate of the true state of the physical system of interest in real-time.

Here, as input vector, use is made of the vectorized density matrix of the cavity state, separated into two parts to account for both real and imaginary components. For target states that do not contain any imaginary part, only the real components are kept to minimize the network dimensions. 
The cavity state is therefore estimated recursively using a quantum filter. Following Haroche's group previous work~\cite{dotsenko_2009, sayrin_2011}, and in absence of decoherence, this filter can be expressed in superoperator form as
\begin{equation}
    \rho_{t+1} =  \mathbf{M}_t\: \mathbf{D}_t\; \rho_{t},
\label{eq:rec_filter}
\end{equation}
with $\rho_t$ being the cavity density matrix estimated at time step $t$, and $\rho_{t+1}$ being the estimate at the next time step $t+1$. $\mathbf{M}_t$ and $\mathbf{D}_t$ are respectively the measurement and displacement superoperators at time step $t$ associated with the measurement operators given in Eqs.~\eqref{eq:meas_ops_g} and~\eqref{eq:meas_op_e} and the displacement operator given in Eq.~\eqref{eq:displacement_op}.

For states with a relative phase, the RL agent outputs two actions, corresponding to the real and imaginary parts of the displacement amplitude $\alpha$ which are limited to be in the interval $\left[-1, +1\right]$. In the case of a state without a relative phase, the agent outputs only one action for the real component of the displacement. The cavity is initialized as an educated guess to a coherent state with mean energy similar to that in the target Fock state superposition. The initial cavity state for the feedback sequence is then:
\begin{equation*}
    \rho_{t=0} =  \mathbf{D}\left(\alpha_\mathrm{guess}\right) \ket{0}\bra{0},
\end{equation*}
where the modulus of $\alpha_\mathrm{guess}$ is given by the mean photon number of the target state. its exact form is then given by $\alpha_{guess} = \sqrt{\overline{n}e^{i\theta}}$ with $\theta$ the relative phase between the superposed states.
After that initial displacement, the agent is then allowed to make an additional one before the measurement sequence in order to fine tune the initial guess.

In the absence of quantum jumps and other decoherence channels, this filtered density matrix corresponds exactly to the cavity state. Below in section~\ref{sec:noise_trajectories}, this filter, along with the cavity state evolution, will be modified to account for noise in the system.

\section{Results}
\label{sec:results}

\subsection{Idealized case}
\label{sec:ideal_sim}

An idealized case is first considered, in which measurements are assumed perfect and the cavity is not subjected to decoherence such as decay and dephasing. Simulations rely on the stochastic evolution of trajectories, updating the cavity state density matrix at every feedback cycle, according to the recursive quantum filter of Eq.~\eqref{eq:rec_filter}. During training, the maximum number of feedback cycles per episode is limited to 50, and the cavity Hilbert space is truncated to $n=29$ photons. Whenever the photon number population in the Fock states $n=28$ or $n=29$ is above a 2$\%$ threshold, the episode is stopped to prevent the RL agent from being biased by Hilbert space truncation.

\begin{figure}[h]
\includegraphics[width = 8.6cm]{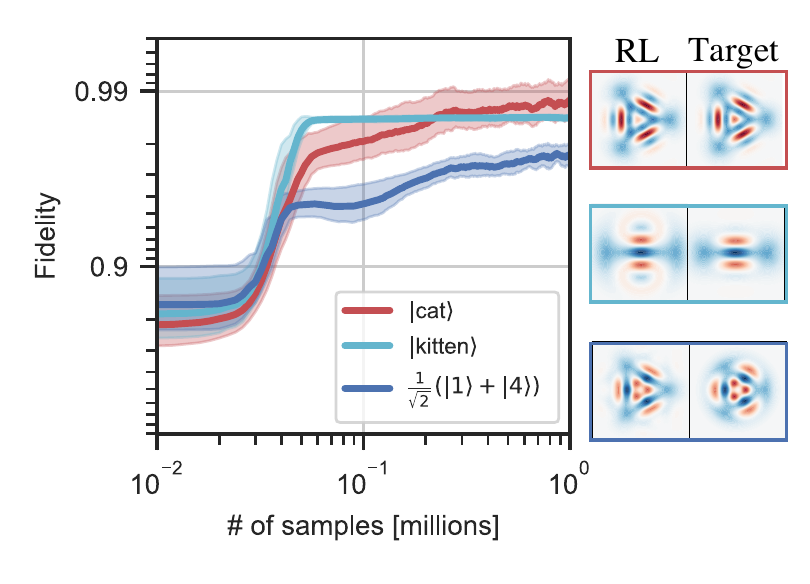}
\caption{Training curves of the TQC RL agent for: i) a 3-components cat state with mean photon number $n = 3$ (red), ii) an equal superposition of the two kitten binomial logical states (light blue), and the state $|\psi\rangle = \frac{1}{\sqrt{2}}\left(|1\rangle + |4\rangle\right)$ (dark blue). On the right are the Wigner functions for prepared states, evaluated after training and compared with the target state.}
\label{fig:training_curves}
\end{figure}

Fig.~\ref{fig:training_curves} depicts the training curves for three different bosonic states, obtained by averaging final states fidelities obtained for 600 trajectories.

The state $|\psi\rangle = \left(|1\rangle + |4\rangle\right)/\sqrt{2}$ is significantly harder to learn for the agent, in part because it has support on only two Fock states. As such, it is harder to control leakage to other states inside the stabilized manifold. This is similar to the limitations of a Lyapunov-based controller, which stops once the state reaches the target subspace, rather than the target state. This state will thus be taken as a benchmark to explore the behavior of the feedback protocol further on. 

\begin{figure}[h]
\includegraphics[width = 8.6cm]{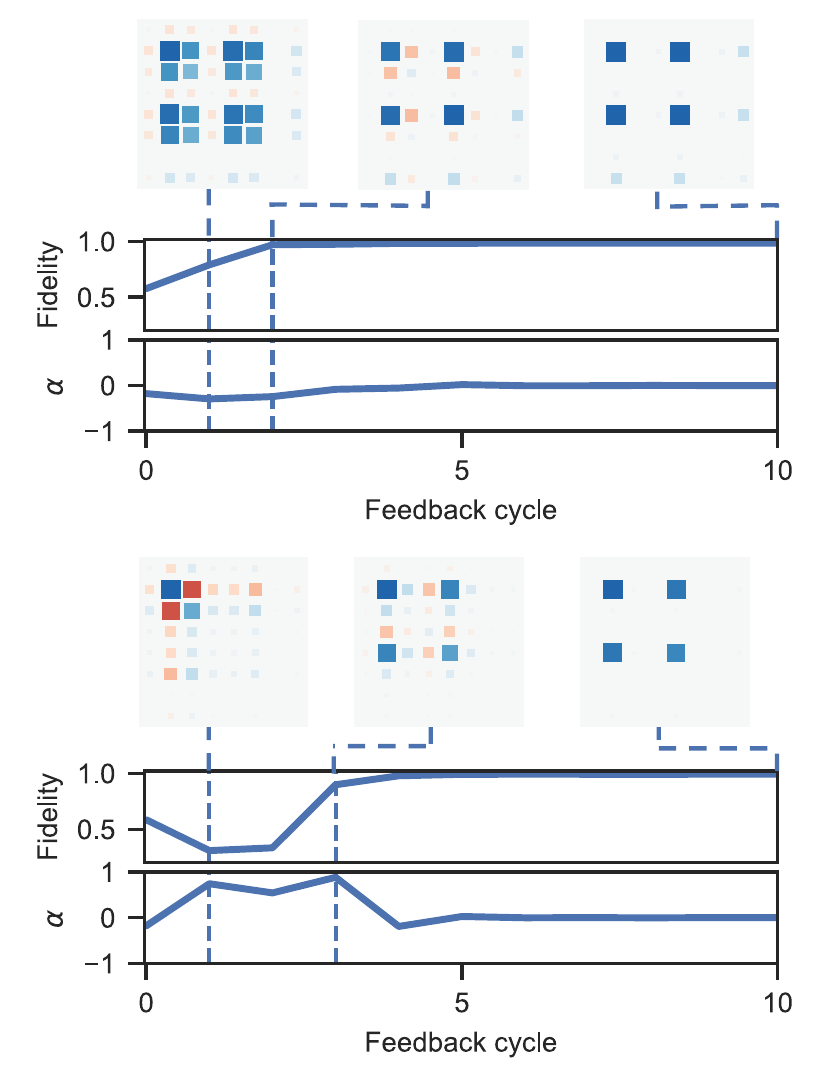}%
\caption{Trajectory examples for the preparation of the state $|\psi\rangle = \left(|1\rangle + |4\rangle\right)/\sqrt{2}$, with Hinton plots of the density matrix at different steps during the preparation sequence. A monotonically increasing fidelity state preparation is shown in the top panel, and the bottom panel shows a trajectory that requires stronger control drives to recover from a sequence of measurements projecting away from the target state.}
\label{fig:sample_curves}
\end{figure}

Fig.~\ref{fig:sample_curves} shows two examples of trajectories that reach the target state $|\psi\rangle = \left(|1\rangle + |4\rangle\right)/\sqrt{2}$, starting from an initial coherent state of amplitude equal to the square root of the target state's mean photon number as mentioned previously. Both cases converge within about 5 feedback cycles. The trajectory in the top panel converges monotonically towards the target state, although with the presence of some leakage to the $\ket{n=7}$ Fock state. The trajectory shown in the bottom panel does not show such leakage, but requires larger displacements to converge to the target state. This illustrates a key benefit of reinforcement learning (RL): As it directly learns the control dynamics from experience, it can handle non-linearities in the control space that are essential for fast convergence, but difficult to handle analytically. 

\begin{figure}[h]
\includegraphics[width = 8.6cm]{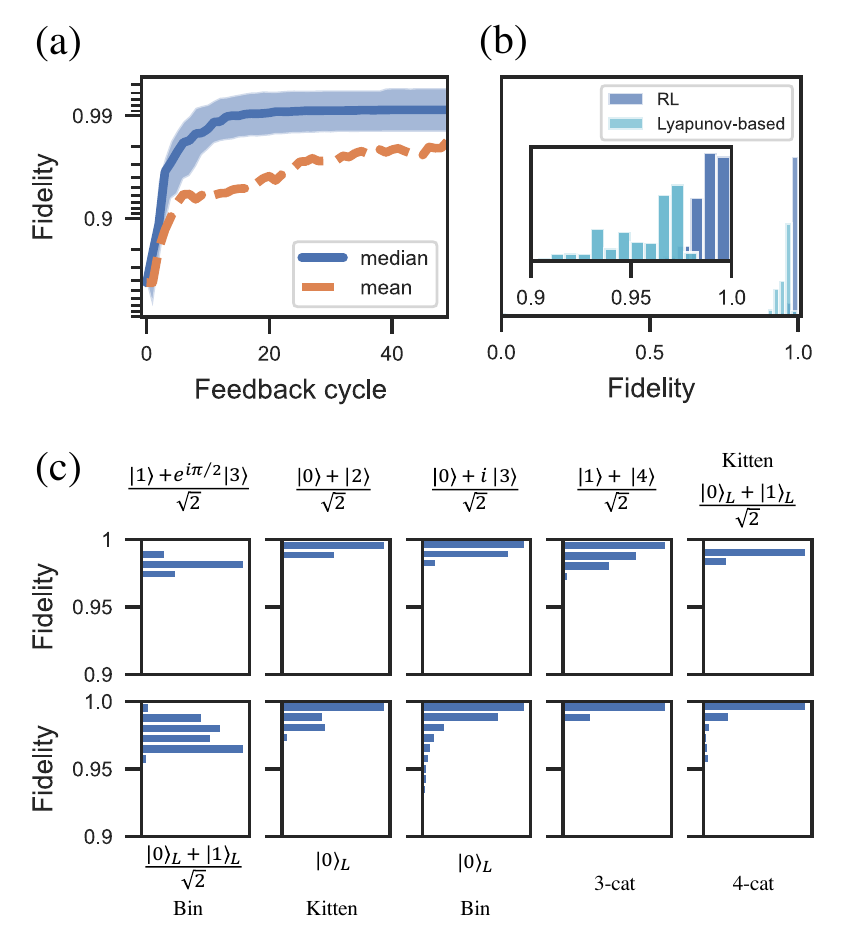}%
\caption{
\textbf{(a)} Time evolution of the state for the preparation of the $(|1\rangle + |4\rangle)/\sqrt{2}$ target state, showing the median with its 25-75 percentile distribution~(shaded area) as a function of time, with the mean converging slower than the median to the target state. Most of the state preparation occurs within the first 10 feedback cycles. \textbf{(b)} Distribution of the final fidelities in (a) for the RL agent alongside a Lyapunov function-based controller. In the inset, it can be observed that the RL procedure converges to higher fidelities, preventing early stopping due to local optima. \textbf{(c)} Final fidelities distribution for a set of different cavity states. All states have most of their distributions above the 95$\%$ mark, even for more difficult binomial encodings. Here, Bin corresponds to the binomial code with support on number states $\{0,3,6,9\}$.
The multi-components cat states are defined on the $n \mod 3 = 0$ number states for the 3-cat state and on $n \mod 4 = 1$ for the 4-cat state.}
\label{fig:bosonics_example}
\end{figure}

Fig.~\ref{fig:bosonics_example}~(a) shows the time-evolution of the fidelity with a fully trained RL agent.
It is seen that the mean fidelity increases slowly compared to the median, Also, 75$\%$ of the trajectories are above 98$\%$ fidelity after about 10 feedback cycles. Comparing this RL agent to a Lyapunov function-based controller (see its derivation in Appendix~\ref{sec:appB}), Fig~\ref{fig:bosonics_example}~(b) shows the fidelity distribution at the end of a 50 cycles sequence. The RL framework outperforms the Lyapunov controller, as its fidelity distribution is clearly above that obtained with Lyapunov control. Remarkably, it also does so without increasing the amount of unsuccessful state preparations. Section~\ref{sec:discussion} further explores the disparities in the two approaches.

Performances of fully trained RL agents for the preparation of a variety of bosonic states are shown in Fig.~\ref{fig:bosonics_example}~(c). In all cases, average fidelities are above 96.6$\%$, with the median consistently above the average by about $2\%$. Similar to the previous case, the distribution is sharply asymmetric, with only a handful of trajectories failing to converge towards the target state after 50 cycles. In a real experiment, as the fidelity of each specific trajectory is tracked with the quantum filter, one could simply discard any trajectory that fails to converge after a given
pre-fixed number of feedback cycles.
Such heralded state preparation has already been realized experimentally~\cite{Liu_2016}. This illustrates a main advantage of measurement-based feedback over open-loop control, and the reason why the median fidelity metric is in this case more representative of the performance than the mean for the proposed scheme.

Unsurprisingly, states with higher mean photon population have lower fidelities, which is likely due to the need for larger displacement drives to prepare these states. These larger drives tend to populate other Fock states within the stabilized subspace. This could be improved by varying the phase shift per photon in time to decimate these in-subspace Fock states, or by directly leaving the measurement parameters as additional controls for the RL agent as in Ref.~\cite{porotti_2022b}.

It was also observed empirically that a smaller $\Delta n$ requires less feedback cycles, caused by the larger sensitivity of the measurements to changes in $m$ (i.e. more efficient discrimination between different subspaces).
This makes the back-action stronger, hence decimating the population in other subspaces faster. Interestingly, superposition states that contain a $\ket{0}$ component have higher fidelities and less dispersion in their distribution. A possible explanation is that the $\ket{0}$ state is a lower bound of the Hilbert space, in the sense that there cannot be negative numbers of photons. Along with the nature of the coherent drive which has a decreasing exponential envelop, this possibly lowers the probability of a transition to another subspace. Such features of the control drive and environment may make it easier for the agent to find the best policy for preparing superpositions.

\subsection{Realistic case with decoherence and imperfections}
\label{sec:noise_trajectories}

Decoherence adds additional loss channels to the system Hamiltonian, with the master equation in Lindblad form now governing the evolution of the state density operator, and given in the rotating frame by
\begin{equation}
    \frac{d\rho}{dt} = \mathbf{L}\rho = \kappa \left(a\rho a^{\dag} - \frac{1}{2}\left(a^{\dag}a \rho + \rho a^{\dag}a  \right)\right) ,
    \label{eq:lindblad_decay}
\end{equation}
where the cavity is assumed to be at zero temperature, so that only photon decay contributes to decoherence. Because of the negligible intrisinc dephasing rate of superconducting cavities, photon loss is therefore the only decoherence channel affecting the cavity considered here~\cite{Reagor_2016}.

Errors coming from the probe qubit are, however, taken into account, as these impact the resulting back-action on the cavity, given by the updated filter terms in Eq.~\eqref{eq:rec_filter} (see also the upcoming Eq.~\eqref{eq:noise_filter} which takes measurement errors into account). 
Since qubit $T_2$ errors commute with the interaction Hamiltonian, they can be considered as occurring after the interaction.
Such errors have an effect that is similar to measurment effors, since both are induced by the $\sigma_z$ operator.
As for qubit $T_1$ errors, which would dephase the cavity, an error transparency method is considered, where relaxation events do not impact the cavity state~\cite{Rosenblum_2018}. So, it is assumed here that our protocol is $T_1$ fault-tolerant.

In summary, qubit errors are modeled in the simulation as an effective $\sigma_{z}$ type error, with the effective decoherence rate being the sum of measurement and $T_{2}$ errors.

\subsubsection{Quantum filter and simulation}

As the agent input should consist of the best possible estimate of the true quantum state of the cavity, the filter update equation (Eq.~\eqref{eq:rec_filter}) needs to be adapted to account for the decoherence channels mentioned above. 

In the simulations, two distinct density matrices are evolved. One evolution concerns the true \textit{cavity} state, where photon loss events correspond to discrete quantum jumps. This is thus a simulation of the true physical system. The other evolution provides a \textit{filtered} estimated state of the cavity which does not have access to these discrete jumps. This filtered estimated state is indeed the only information available about the actual state in practice for control purposes (built from measurements of the true system), since having full information about the state is impossible. The latter evolution estimates the cavity relaxation using a first order expansion of the Lindblad dissipator given in Eq.~\eqref{eq:lindblad_decay}, which can be added to the filtering equation with a superoperator of the form
\begin{equation}
    \mathbf{T}\rho = \left(\mathbf{1} + T_\mathrm{cav} \mathbf{L}\right)\rho,
\end{equation}
where $T_\mathrm{cav}$ is the cavity lifetime. In the case of a discrete jump, the filter will then move away from the actual cavity state, while subsequent measurements obtained from the true cavity simulation will update the filter until it converges back towards the real cavity state. This is the beauty of quantum filtering, which is analogous to Kalman filtering in classical control theory, as it is able to improve the estimate of the state as information is accumulated over time, consisting of the known measurements results obtained and the actions performed on the system.

Due to the finite measurement accuracy of the probe 
qubit, the resulting estimated state is a statistical mixture of the two possible outcomes. This is taken into account with the following superoperators
\begin{align}
    \mathbf{P}_e \rho &= \left(1- P_{f,e}\right) \mathbf{M}_e\rho + P_{f,e}\mathbf{M}_g \rho , \\
    \mathbf{P}_g \rho &= \left(1- P_{f,g}\right) \mathbf{M}_g\rho + P_{f,g}\mathbf{M}_e \rho ,
\end{align}
where the weights $P_{f,e}$ and $P_{f,g}$ of each measurement operator depend on the erroneous state assignation probabilities $\eta_{e|g}$ and $\eta_{g|e}$ in a measurement. Following Ref.~\cite{sayrin_2011}, in the case of a measurement that delivers $e$, whereas it should have been $g$, the weight is given by $P_{f,g} = \eta_{e|g}P_e / [\left(1- \eta_{e|g}\right)P_g + \eta_{e|g}P_e ]$, where $P_e = \tr (M_e \rho M_e^{\dagger})$ and $P_g = \tr (M_g \rho M_g^{\dagger})$. In the other case that a measurement delivers $g$, whereas it should have been $e$, the weight is $P_{f,e} = \eta_{g|e}P_g / [\left(1- \eta_{g|e}\right)P_e + \eta_{g|e}P_g ]$.
The estimated state is then updated using the following modified recursive filtering equation
\begin{equation}
    \rho_{t+1} =  \mathbf{P}_{t}\: \mathbf{T}\; \mathbf{D}_{t}\; \rho_{t}.
\label{eq:noise_filter}
\end{equation}
It is to be noted that only the filtered state is given as input to the agent, as this would be the only information available in a real experiment.

\subsubsection{Simulation results}

Following a procedure similar to that in section~\ref{sec:ideal_sim}, 3000 trajectories are simulated, each consisting of 2000 feedback cycles (the smaller number of total trajectories simulated here compared to than in section~\ref{sec:ideal_sim} is due to the need to simulate more feedback cycles for each trajectory in the present case). In these simulations, a time of 1~$\mu$s is considered for one feedback cycle (300~ns for the cavity-qubit interaction time, while leaving 700~ns for qubit measurement and rotations; according to literature, these values are well within current device performances~\cite{heeres_2017,Krinner_2022}). The cavity lifetime considered here is 1~ms~\cite{Reagor_2016,Sun2014}.
The errors in qubit state measurement assignments are $\eta_{e|g} = 0.01$ and $\eta_{g|e} = 0.02$.

\begin{figure}[ht!]
\includegraphics[width = 8.6cm]{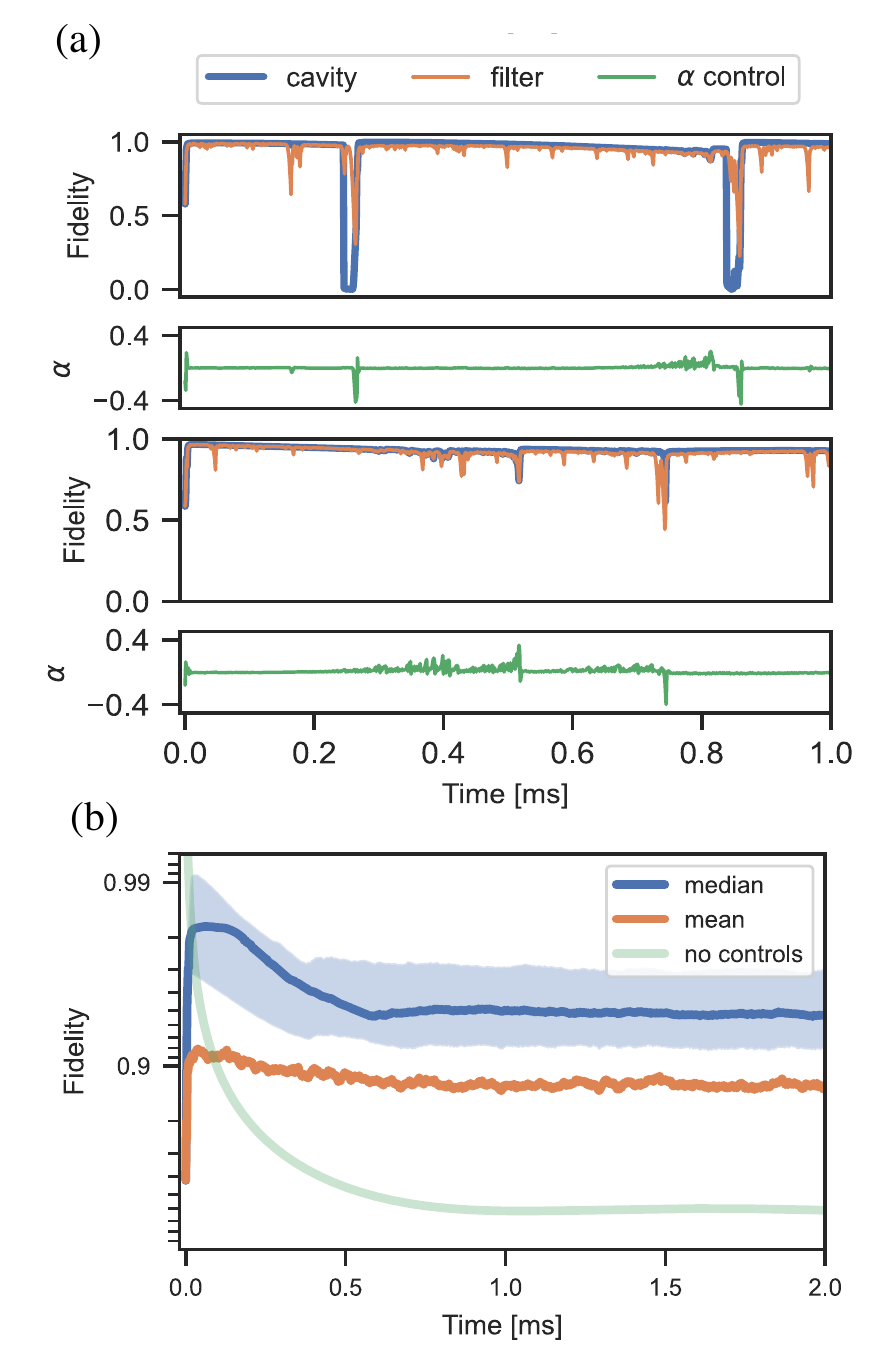}%
\caption{Preparation and stabilization of the $\big( |1\rangle +|4\rangle \big) / \sqrt{2}$ state under decoherence. \textbf{(a)}  The RL agent is able to recover from photon loss events, as shown in top panel.
When no photon loss occurs (bottom panel), Zeno dynamics take place, with population transfer occuring only inside the target subspace (due to measurement back-action suppression of population leakage out of the subspace), but which does not prevent the slow decay of the fidelity.
The control amplitude $\alpha$ (green curve) is 0 for most of the sequence, except when the filter (orange curve) detects a photon loss, or when the fidelity goes below certain value (bottom panel). \textbf{(b)} Time evolution for a cavity lifetime $T_1$ of 1~ms and a feedback cycle time of 1~$\mu$s. Median fidelities are still able to reach values similar to the ideal case, although they stabilize around about 95$\%$ when multiple photon loss events occur. The mean fidelity remains at 90$\%$ throughout, pushed down by the momentary photon loss. With perfect initialization at $t=0$, the average state fidelity would have decayed according to the master equation, as shown by the green curve.}
\label{fig:loss}
\end{figure}

Fig.~\ref{fig:loss} shows state evolution and preparation results with decoherence, with~\ref{fig:loss}~(a) depicting trajectory examples with noticeable photon loss events and subsequent recovery. In this case, the agent is able to recover from photon loss events with a delay depending on the speed at which the filter recognizes the loss event. An interesting behavior is shown in the bottom panel of Fig.~\ref{fig:loss}~(a), where no photon loss events are registered; the cavity state instead evolves deterministically inside the stabilized subspace as indicated by the small and slow decay of the fidelity between jumps. This decay is due to the slow population transfer from the $|4\rangle$ state to the $|1\rangle$ state; this comes from the Zeno dynamics induced by the back-action of the measurements~\cite{Facchi_2002}, whereby the evolution of the state is slowed down by the measurements. The out-of-subspace leakage from the deterministic evolution of the stochastic master equation between quantum jumps is hence being suppressed by the measurements. To correct this slow fidelity decay, the RL agent has limited control. Indeed, between jumps, the controller shows a jittering behavior of growing amplitude lasting over 0.1~ms as the fidelity decays. This is distinct from the behavior following a photon loss, where the correction applied appears as an isolated sharp peak in the control amplitude. This phenomenon will be discussed further in section~\ref{sec:discussion}, where it will be shown that small corrections once near the target state are not the primary means to achieve high fidelity.
Fig.~\ref{fig:loss}~(b) shows the behavior of an ensemble of trajectories, for the same loss parameters as above. Unsurprisingly, the mean is significantly lower, as photon loss events, occurring randomly, drag it downwards. The median stabilizes around a fidelity of 95.5$\%$, after having reached fidelities of 98$\%$. This drop can be explained by the RL agent failing to recover fully from photon losses on average. 

To compare with the situation corresponding to no control (free evolution), the green curve shows the average evolution where perfect state preparation is assumed at time $t=0$. This shows that the approach proposed here, even with limited control, namely measurement back-action combined with coherent driving in a feedback loop, helps stabilize the target state.

\section{Discussion}
\label{sec:discussion}

\subsection{Robustness to noise}

\begin{figure}[h]
\includegraphics[width = 7.5cm]{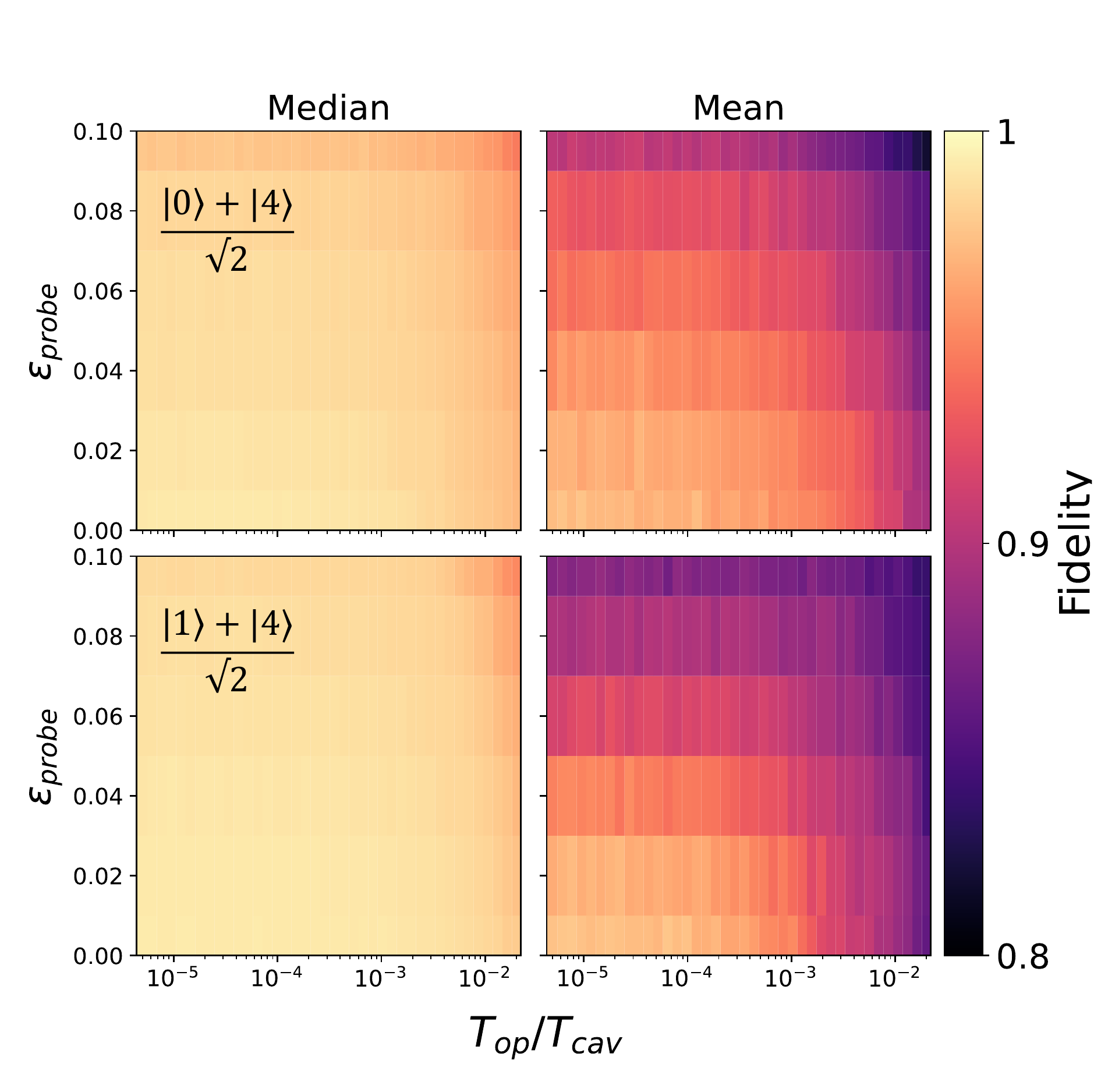}%
\caption{Maximum fidelities reached by the RL agent during state preparation as a function of decoherence parameters, with the horizontal and vertical axes corresponding respectively to the cavity lifetime and the probe 
qubit's errors.
}
\label{fig:noise_sweep}
\end{figure}

To further analyze the robustness of the TQC agent proposed here, its performance is studied with a fully trained agent for ranges
of cavity lifetimes and probe qubit imperfections.
Note that all results were obtained using the same RL agent, trained on the ideal model. It was found that such a model usually performs better than one trained on a lossy system. One possible explanation is that the Markovian nature of the quantum filter combined with the exploration properties of the RL training are sufficient to learn an optimal policy. Indeed, as the RL agent learns about the environment, it explores states similar to those resulting from quantum jumps. However, contrary to the case with significant decoherence, it is also able to explore high fidelity states, which then makes it a more complete agent, able to perform well under different system dynamics.
In Fig.~\ref{fig:noise_sweep}, the lifetimes are expressed as the ratio of the total feedback cycle operation time (1~$\mu$s is considered here) and the cavity lifetime (which is varied to include currently experimentally realistic values).
Also, as mentioned in section ~\ref{sec:noise_trajectories}, errors from the probe 
qubit are summarized into an effective error denoted $\epsilon_\mathrm{probe}$, consisting of the sum of all individual qubit decoherence channels. Fig.~\ref{fig:noise_sweep} shows the maximum median and average fidelities attained during a 50 cycles state preparation procedure, as a function of both cavity decay and qubit errors for the $\big(|1\rangle + |4\rangle\big)/\sqrt{2}$ and $\big(|0\rangle + |4\rangle\big)/\sqrt{2}$ states, which respectively have an odd and even $\Delta n$.

For the median fidelity metric, it is seen in Fig.~\ref{fig:noise_sweep} that both states are robust to a large range of parameters, with a drop in fidelity being seen in the top right corner corresponding to high error rates in both cavity and qubit. The impact of decoherence is unsurprisingly more pronounced for the mean fidelity metric, indicating that while the majority of state preparation sequences lead to high fidelity states, a small fraction, however, completely fail to reach high fidelity. 

The drop in the median fidelity value as measurement errors are more prominent can be attributed to a state estimation problem for the controller. When measurement errors are high, the quantum filter needs more measurement results to construct an accurate estimate of the true cavity state. On some occasions, the displacement applied to the cavity can increase the deviation between the estimated state and the true cavity state in such a way that the former fails to converge back to the cavity state. Such instances of deviation between the cavity and the estimated states happen more frequently following a photon loss, which could also explain why the state $\big(|1\rangle + |4\rangle\big)/\sqrt{2}$ is more subject to decoherence, as it has a slightly larger mean photon number, and also contains the state $|1\rangle$, which can still decay down to $|0\rangle$ as compared to the state $\big(|0\rangle + |4\rangle\big)/\sqrt{2}$ in which $|0\rangle$ cannot further decay down. 

\subsection{Understanding the trained policies}

Attention will now be turned to the policies learned by different types of agents. The TQC agent proposed herein will be compared with two other approaches: the RL-based PPO and the Lyapunov based controller mentioned previously. This controller chooses the best displacement drive by performing a line search over a linearization of the Lyapunov fidelity following application of the displacement operator to the current cavity state, see Appendix~\ref{sec:appB}.

\begin{figure}
\includegraphics[width = 8cm]{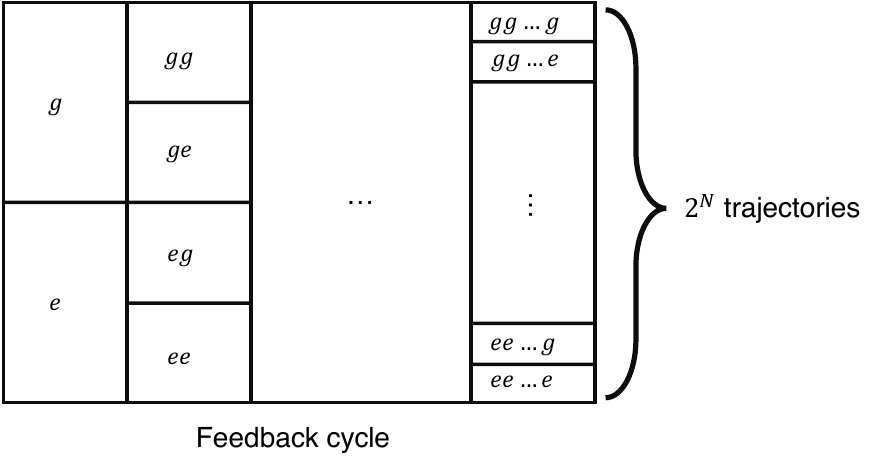}%
\caption{Structure of the trajectories calculations presented in Fig.~\ref{fig:policies}. At each additional feedback cycle, the binary tree is expanded by a factor of two, corresponding by two possible measurements outcomes in each branch.}
\label{fig:trajectory_grid}
\end{figure}

\begin{figure*}
\includegraphics[width = 17cm]{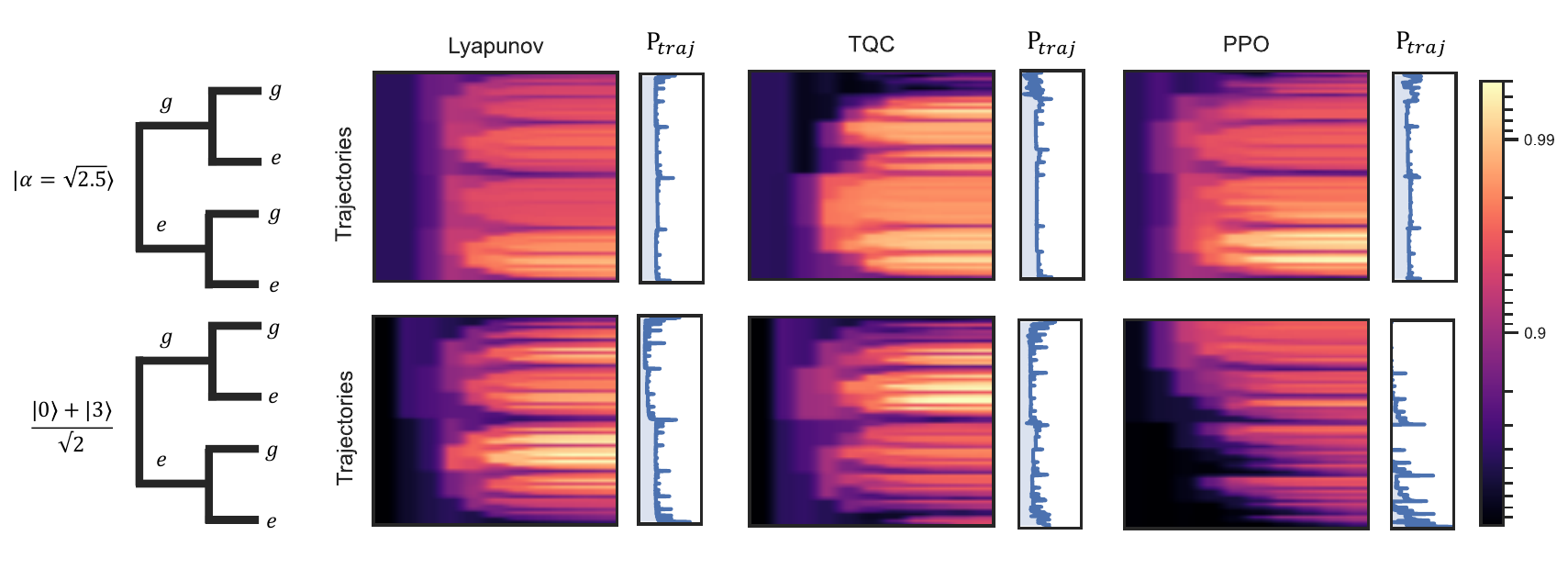}%
\caption{Comparison of policies learned by different agents to prepare the state $\frac{\ket{1} + \ket{4}}{\sqrt{2}}$. Top row left: Schematic of the exhaustive trajectory search in the form of a binary tree. The initial state is a coherent state with a mean photon number of 2.5; each branch corresponds to a specific trajectory outcome. The rest of the top row shows the evolution of the fidelity along each of the $2^{10}$ trajectories, for Lyapunov-based control, the TQC and PPO RL agents. Also shown are the probabilities of occurence of the different trajectories plotted on a log scale. RL based methods perform best, owing to their strong initial displacements allowing to reach higher final fidelities. Bottom row:  Same as in top row, but initializing with the state $\frac{\ket{0} +\ket{3}}{\sqrt{2}}$, which is in a stabilizable subspace different than that of the target state. Lyapunov and TQC agents are both able to reach the target subspace. The PPO agent fails at learning any policy, which indicates a lack of exploration during training.}
\label{fig:policies}
\end{figure*}

The policy space has the form of a binary tree with $2^N$ possible trajectories as each measurement leads to two distinct possibilities in the decision tree. Here, $N$ is the depth of the tree corresponding to the number of feedback cycles, which in this case is chosen to be 10 as a compromise so that the computation time does not become prohibitive since here all trajectories are exhaustively examined.
Fig.~\ref{fig:trajectory_grid} provides details on the procedure and how the results in Fig.~\ref{fig:policies} are obtained. Starting from a given initial state, the cavity states of all possible combinations of $g$ and $e$ measurement outcomes are computed at every feedback cycle. The corresponding metric, that is the fidelity between the cavity state and the target state, is given by
\begin{equation}
F\big( M_{s_{k}}D_{s_{k}},...,M_{s_{2}}D_{s_{2}},M_{s_{1}}D_{s_{1}}\ket{\psi_{0}} , \ket{\psi^\mathrm{target}} \big). 
\end{equation}
To each measurement outcome corresponds a probability of occurrence, with the product of these individual probabilities for a given trajectory being the probability of occurrence of the whole trajectory.

Fig.~\ref{fig:policies} shows such trajectory distributions, with the color scale corresponding to the fidelity after a measurement at a given timestep (feedback cycle), lighter colors corresponding to higher fidelities. Displaying the fidelity in this manner provides a high-level view of the policies learned, and tells how far from the target state (and thus out of subspace) the agent can go in order to maximize the final fidelity.

\subsubsection{Starting from a coherent initial state}

Trajectories are shown in the top row of Fig.~\ref{fig:policies}, with the initial state of the feedback sequence being the same initial coherent state used during training of the RL agents.
There are differences between the agents in the structure of their trajectories. For instance, the Lyapunov controller has well defined paths in the fidelity space, influenced by previous measurements results. This behavior is similar for the PPO agent, although less pronounced. These branchings are much less apparent for the TQC agent. This indicates a better ability to exploit the combined effect of measurement back-action and coherent drive. In other words, RL agents, and especially the TQC agent, appear to be less influenced by the measurement outcomes than they are exploiting them.
It is seen in Fig.~\ref{fig:policies}, especially in the upper part corresponding to a first ground state measurement, that both RL agents learn to do penalizing displacements in the early feedback cycles, that will however allow reaching higher fidelities later on, and significantly more so for the TQC agent. Although the first few timesteps have lower fidelities compared to the other two approaches, the TQC agent is nevertheless able to reach high fidelity states early on in the control sequence.

\subsubsection{First feedback cycles}

\begin{figure*}
\includegraphics[width = 17cm]{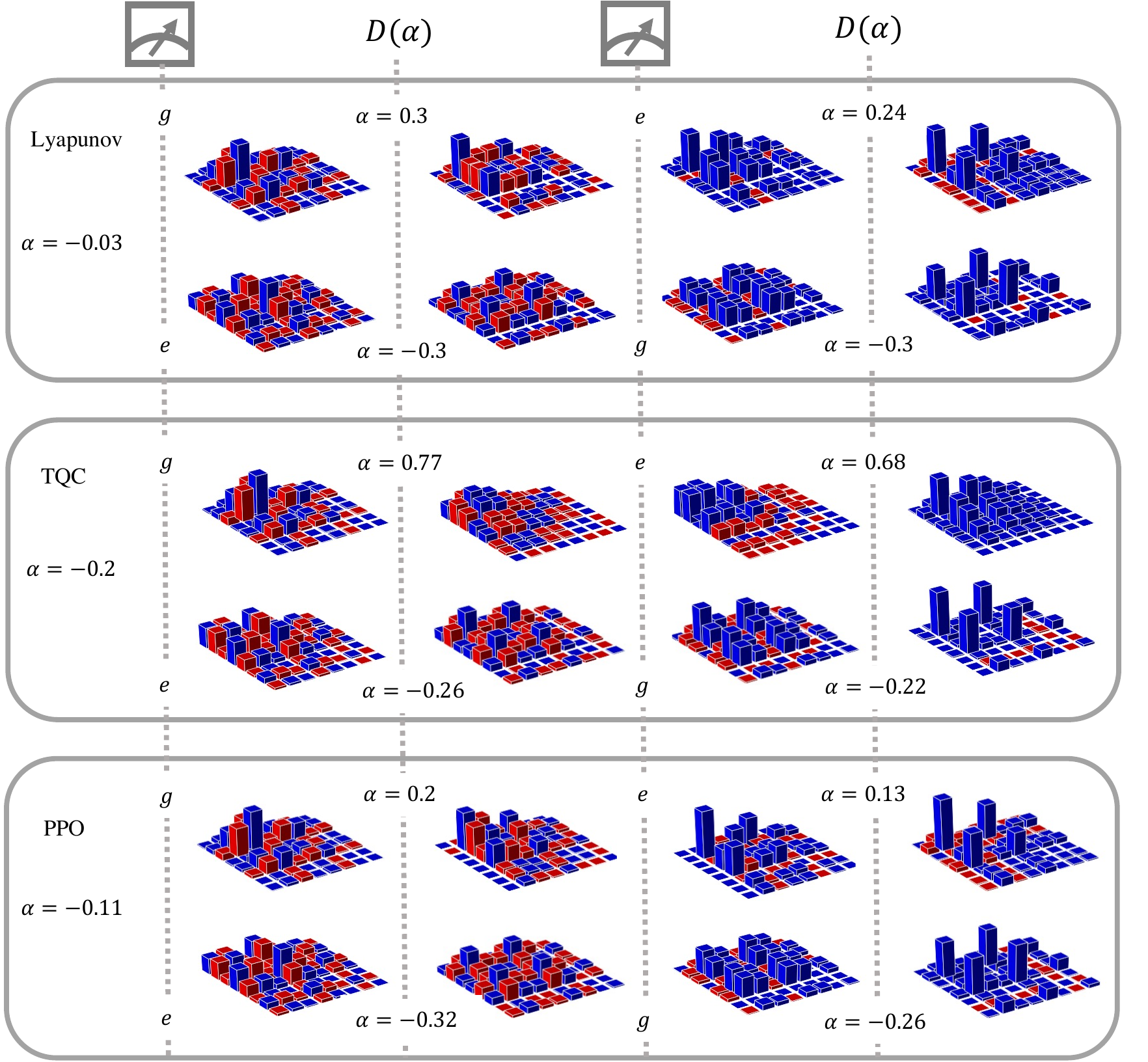}%
\caption{First feedback cycles for the cavity initialized as a coherent state, showing how the cavity states evolves at each step of measurements and displacements. For each agent, two different trajectories are shown, consisting of a sequence of g-e or e-g measurement outcomes (respectively top and bottom row). The $\alpha$ value at the left before the first measurement corresponds to the first control, adjusting the initial $\alpha_{guess}$ value. While the bottom row is similar for all agents, it is seen in the top row that the TQC agent opts for a drastically different strategy by performing a large displacement to reset the cavity in a state similar to a coherent state.}
\label{fig:first_feedback}
\end{figure*}

The aforementioned behavior of the first few feedback cycles is depicted
in Fig.~\ref{fig:first_feedback}, where the evolution of the cavity density matrix is shown for all approaches.

In situations where the current state is far from the target, a Newton-based line-search method, such as that used in the Lyapunov control approach implemented here (see Appendix~\ref{sec:appB}), will not optimally determine the displacement needed to be applied to the cavity state.
This is a consequence of the locally convergent behavior of line-search methods. Indeed, it was found that the Lyapunov control was constantly selecting large displacements in such cases, which prevented further convergence to high fidelity states. This required to limit the maximum allowed displacement for the Lyapunov control. This was here done by optimizing over a range of candidate maximum amplitudes, and choosing the one maximizing the median fidelity to the target state.
The $\alpha = \pm 0.3$ displacement performed by the Lyapunov control approach shown in Fig.~\ref{fig:first_feedback} is the result of such optimization. Compared to the RL methods, this a serious drawback as the choice of the large displacement to be applied early in the control sequence is crucial to adequately balance the amplitudes of the resulting state after the control sequence, and should ideally be conditioned on a specific state, rather than be optimized over all trajectories.

For RL agents, it is indeed found that they select displacements of similar amplitudes applied to the cavity after the first excited state qubit measurement, but which is better adapted as it takes into account the impact it may have in future feedback cycles. This is even more striking in the case following a ground state measurement at the beginning of the sequence. Here, the TQC agent performs a large displacement of amplitude 0.77, which effectively does an operation akin to a state reset. On a short time scale, this is penalizing as shown in the states that follow in the sequence, which are farther from the target state than for the other approaches. However, as seen in the top row of Fig.~\ref{fig:policies}, this opens the way to states with high fidelity, and indeed higher fidelities are reached than with the other approaches. The TQC agent is thus better at exploiting the effect of the measurement back-action to reach its target state. 

One can also notice the larger initial displacement, in absolute value, taken by the RL agents, correcting the initial $\alpha_\mathrm{guess}$ value. RL agents are able to infer, simply by retro-propagating future fidelities observed, that the initial state, even though it is one with the best overlap with the target state, is not the optimal one when taking into account the measurement dynamics, and when considering the global objective to be reached on a longer horizon.

\subsubsection{Out of subspace initial state}

Finally, the bottom row of Fig.~\ref{fig:policies} shows results of a sequence initialized with the state~$|\Psi\rangle =\left(|0\rangle + |3\rangle\right)/\sqrt{2}$, which has amplitudes similar to those of the target state, but which is in another subspace. This case is perhaps the most interesting, as the different approaches show drastically different behaviors. PPO is not able to learn how to leave the subspace, as shown by the higher probabilities for trajectories that are associated to low fidelity states. Indeed, it was found that it only performs small displacements of about $\alpha = 0.03$ which are not sufficient to transfer enough population to the target subspace, with the back-action simply annihilating all target subspace populations at every step.
In this situation, Lyapunov control performs better. In this case, it applies the maximum allowed displacement $\alpha = 0.3$ at the beginning (this cannot be seen from the figure; it is the value given by the algorithm), which allows it to eventually reach the target subspace. The TQC agent, however, appears to find the best policy ($\alpha = 0.54$ is applied at the first cycle of feedback - value from the algorithm), transferring 40$\%$ of subsequent trajectories towards high fidelity values. It is also able to learn how to bring a large portion of the remaining 60$\%$ towards higher fidelities. In fact, as the TQC policy evolved during training, it was seen to opt for a trade-off. In early stages, it was performing a larger initial displacement, which favored a higher probability for a measurement to bring it back to the target subspace. However, it did so at the expense of the remaining trajectories, which never reached the target state. As training evolved, it lowered the initial displacement, so as to still handle the remaining states resulting from the other measurement outcomes. It should also be emphasized that during training, the agent was never initialized in the $\left(|0\rangle + |3\rangle\right)/\sqrt{2}$ state, but only in the coherent state as mentioned above. As such, the policy found by the TQC agent is a result of its higher exploration and generalization capabilities.

This analysis shows that the simple extension of the task of preparing a single Fock state to that of preparing superpositions translates into a qualitatively different control problem. Whereas the former only needs the state to be steered towards the desired eigenstate of suitably chosen measurement operators, the latter requires a complex interaction between measurement outcomes and displacement operations.  In some cases, even something akin to a state reset is necessary in order to maximize the achievable fidelity.

It should also be noted that while an on-policy was compared to an off-policy method here, it is still unclear what is the role of adding a distributional approach on top of a soft-actor critic algorithm.
Nevertheless, compared to the standard soft actor-critic implementation~\cite{haarnoja_2018}, it was found in the present work that the distributional version is more robust to hyperparameters tuning and more stable during training. More work would however be needed, such as ablation studies, to better understand the role of the distributional procedure in the quantum setting.

\section{Conclusion}

In this work, a measurement-based quantum-feedback protocol was proposed and analyzed to prepare and stabilize superpositions of Fock states in a superconducting cavity. By using a generalization of parity measurements, states in a target subspace with Fock basis states with equally spaced number of photons can be stabilized, but also prepared using a coherent drive as the only control. It was shown that a classical control technique such as Lyapunov function-based control, which was previously developped for the stabilization of Fock states~\cite{dotsenko_2009,sayrin_2011}, fail to prepare superposition due to their lack of exploitation of the measurement back-action. 

Here, using an RL method proved useful in overcoming this limitation. Indeed, by exploring the optimal policies learned by different algorithmic methods, the measurement back-action could be emphasized as a useful resource in itself to create the non-linearities required to prepare quantum state superpositions. It also highlighted the interaction between control actions with weak measurements in a way that takes into account the impact of the control actions later in the feedback cycle, in order to reach a target state with high fidelity.
Because such measurements are the same as those
 used in some error-correction procedures, our proposed protocol could easily be integrated in the bosonic computation paradigm.

From an RL point of view, the ability of TQC to reuse past experiences, and thus learn a more general policy than, for instance, PPO-like algorithms, might prove useful as future quantum control experiments scale up in complexity. We believe that further exploration of these different behaviors already noticed in the low complexity settings considered here would be an interesting and potentially fruitful avenue of investigation.

\begin{acknowledgements}
This project was supported by Institut quantique (IQ) at Université de Sherbrooke (UdS) through the Canada First Research Excellence Fund. AP acknowledges support from the QSciTech program funded through an NSERC-CREATE grant. YBL acknowledges fruitful discussions with Pierre Rouchon and Rémi Azouit.
\end{acknowledgements}

\appendix
\section{Neural networks hyperparameters}
\label{sec:appA}

Tables~\ref{table:HyperparamsTQC} and~\ref{table:HyperparamsPPO} give the hyperparameter values used for both the TQC and PPO agents. All others hyperparameters not presented here are those by default in the Stable-Baselines~3 implementation.

\begin{table}[h]
\caption{Hyperparameters for TQC.}
\label{table:HyperparamsTQC}
\centering
\begin{tabular}{|c|c|} 
\hline
\textbf{Hyperparameter} & \textbf{Value} \\ [0.5ex] 
\hline\hline
Number of Layers & 2 \\ 
\hline
Actor Neurons per Layer & 256 \\
\hline
Critics Neurons per Layer & 512 \\
\hline
Discount($\gamma$) & 0.95 \\
\hline
Batch size & 1024 \\
\hline
Activation Function & tanh \\
\hline
Entropy coeffcient & 0.09 \\
\hline
Number of critics & 5 \\
\hline
Learning Rate & 0.0001 \\
\hline
Target update interval ($\tau$)  & 0.001  \\ [1ex] 
\hline
\end{tabular}
\end{table}

\begin{table}[h]
\caption{Hyperparameters for PPO.}
\label{table:HyperparamsPPO}
\centering
\begin{tabular}{|c|c|}
\hline
\textbf{Hyperparameter} & \textbf{Value} \\ [0.5ex] 
\hline\hline
Number of Layers & 2 \\ 
\hline
 Neurons per Layer & 256 \\
 \hline
Discount($\gamma$) & 0.95 \\
\hline
 Number of steps & 2048 \\
\hline
Batch size & 256 \\
\hline
Activation Function & tanh \\
\hline
Learning Rate & 0.0001 \\ [1ex] 
\hline
\end{tabular}
\end{table}

\section{Derivations for Lyapunov function-based control}
\label{sec:appB}

In Lyapunov function-based control, a positive-definite function $V$ of the state $\rho$ is used, so that the minimum of $V$, which is necessarily zero owing to positive-definiteness, is reached for a targeted state $\rho^\mathrm{target}$. Control actions are performed iteratively in a feedback loop, whereby the goal of each iteration's control action on $\rho$ is to reduce the value of $V$, ultimately reaching the minimum, hence the targeted state~\cite{mirrahimi_stabilizing_2007}.
The essential details of this approach in the present specific context will now be provided.

\subsection{Positive-definite and Lyapunov functions and their significance in control}

Given two arbitrary states $\rho_1$ and $\rho_2$, a positive-definite function $d(\rho_1, \rho_2)$ will be considered in the sequel, that is with the following properties
\begin{equation}
  d(\rho_1, \rho_2)
  \left\{\begin{array}{l} > 0 \,\,\,\, \text{for} \,\,\, \rho_1 \neq \rho_2 , \\
                        = 0 \,\,\, \Leftrightarrow \,\, \rho_1 = \rho_2 .
         \end{array}\right.
\label{eq:CondPosDefFct2States}
\end{equation}
This is akin to a distance function, but it is not required here that $d$ obeys the triangle inequality (which is one of the axioms that a distance function must satisfy).
Given such a function and a targeted state $\rho^\mathrm{target}$, one can in turn define a positive-definite function over the set of states $\rho$ by
\begin{equation}
  V^\mathrm{target}(\rho) = d(\rho^\mathrm{target}, \rho) .
\end{equation}
The significance of such a function in the present context is that if $\alpha$ in Eq.~\eqref{eq:CohDriveDensMatSuperopForm} is chosen so that
\begin{equation}
  V^\mathrm{target}(\rho') < V^\mathrm{target}(\rho) ,
\end{equation}
then by performing a series of steps $k = 1, 2, \ldots$, whereby at each step this inequality is satisfied,
\textit{i.e.}
$V^\mathrm{target}(\rho_{k+1}) < V^\mathrm{target}(\rho_k)$, then the state will converge to the targeted state since the value of $V^\mathrm{target}(\rho_k)$ will eventually reach zero, and, by hypothesis, for $V^\mathrm{target}(\rho_k)$ to equal zero the only possibility is that $\rho_k = \rho^\mathrm{target}$.
Such a function decreasing over the evolution of a system is known in the
control litterature as a Lyapunov function~\cite{khalil_2001}.

\subsection{Fidelity-based positive-definite and Lyapunov functions}

One of the simplest positive-definite functions that can be considered in quantum mechanics is based on the \textit{Frobenius scalar product} between two operators, which for density operators is given by
\begin{equation}
  Fr(\rho_1, \rho_2) = \tr(\rho_1 \rho_2) ,
\end{equation}
which for pure states $\rho_1 = \ket{\psi_1}\bra{\psi_1}$ and $\rho_2 = \ket{\psi_2}\bra{\psi_2}$ amounts to
$Fr(\rho_1, \rho_2) = \abs{ \braket{\psi_1}{\psi_2} }^2$.
The Frobenius scalar product is also sometimes simply called the fidelity~\cite{dotsenko_2009} (note that this fidelity is different from that defined in Eq.~\eqref{eq:FidelitySqrRoots}).
Since $0 \leq Fr(\rho_1, \rho_2) \leq 1$ with $Fr(\rho_1, \rho_2) = 1 \,\,\, \Leftrightarrow \,\, \rho_1 = \rho_2$ and $\rho_1$ is pure, one can define the positive-definite function
\begin{equation}
  d_{Fr}(\rho_1, \rho_2) = 1 - Fr(\rho_1, \rho_2) ,
\end{equation}
which satisfies the conditions given in Eq.~\eqref{eq:CondPosDefFct2States}. This will be called the \textit{Frobenius distance}, although it does not formally satisfies all the axioms of a distance.
This leads to the positive-definite function defined over states
\begin{equation}
  V_{Fr}^\mathrm{target}(\rho)
    = 1 - Fr(\rho^\mathrm{target}, \rho)
    = 1 - \tr( \rho^\mathrm{target} \rho ) .
\end{equation}
Since $\tr \rho = 1$, this can be written as
\begin{equation}
  V_{Fr}^\mathrm{target}(\rho)
    = \tr\left( (\mathbb{I} - \rho^\mathrm{target}) \rho \right)
    = \tr\left( \Upsilon^\mathrm{target} \rho \right) ,
\label{eq:LyapunovFidel}
\end{equation}
where $\mathbb{I}$ is the identity operator, and $\Upsilon^\mathrm{target} = \mathbb{I} - \rho^\mathrm{target}$.
Here the superscript "$\mathrm{target}$" on $\Upsilon$ reminds that $\Upsilon$ depends on the targeted state.
Such a Lyapunov function will be called a \textit{fidelity Lyapunov function}, and will be denoted $V_\Upsilon^\mathrm{target}(\rho)$ in the sequel, hence
%
\begin{equation}
  V_\Upsilon^\mathrm{target}(\rho)
    = \tr\left( \Upsilon^\mathrm{target} \rho \right) .
\label{eq:GeneralizedLyapunovFct}
\end{equation}
%

\subsection{Second order expansion of the Lyapunov function with actuator action}

It will now be assumed that a Lyapunov function given in the generalized form of Eq.~\eqref{eq:GeneralizedLyapunovFct} is defined. The objective of the feedback law can now be stated by requiring that, for a given state $\rho$, $\alpha$ must be chosen such that
$
%
  V_\Upsilon^\mathrm{target}(\rho') < V_\Upsilon^\mathrm{target}(\rho) ,
$
that is
\begin{equation}
  V_\Upsilon^\mathrm{target}\left( {D}(\alpha) \rho \, {D}(-\alpha) \right)
    < V_\Upsilon^\mathrm{target}(\rho) .
\label{eq:IneqChoiceAlpha}
\end{equation}
Obtaining such a condition is similar to a line search in numerical minimization~\cite{Fletcher_2000} and a common way is to develop to second order in $\alpha$ the left hand side of this inequality, and use the approximation thus obtained to find a value of $\alpha$ that will allow satisfying this condition. To do this, Eq.~\eqref{eq:DispCavState} is first expanded to second order in $\alpha$ by resorting to an expansion of ${D}(\alpha)$ to second order. This leads to
\begin{align}
  \rho'
    &= {D}(\alpha) \rho \, {D}(-\alpha) \notag \\
    &\approx
       \rho + [\alpha a^\dagger - \alpha^* a, \rho]
       + \frac{1}{2}
         \left[ [\rho,  \alpha a^\dagger - \alpha^* a], \alpha a^\dagger - \alpha^* a \right] .
\end{align}
With this, $V_\Upsilon^\mathrm{target}(\rho')$ can be approximated to second order by
\begin{align}
  V_\Upsilon^\mathrm{target}(\rho')
    &=
      \tr\left( \Upsilon^\mathrm{target} \rho' \right) \notag \\
    &\approx
      V_\Upsilon^\mathrm{target}(\rho) + T^{(1)}(\alpha) + \frac{1}{2} T^{(2)}(\alpha) ,
\label{eq:LyapunovFct2ndOrder}
\end{align}
where $T^{(1)}(\alpha)$ is the first order term given by
\begin{equation}
  T^{(1)}(\alpha) = \tr\left( \Upsilon^\mathrm{target} [\alpha a^\dagger - \alpha^* a, \rho] \right) ,
\end{equation}
and $T^{(2)}(\alpha)$ is the second order term given by
\begin{equation}
  T^{(2)}(\alpha)
    = \tr\left( \Upsilon^\mathrm{target}
                    \left[ [\rho,  \alpha a^\dagger - \alpha^* a],
                             \alpha a^\dagger - \alpha^* a \right]
          \right) .
\end{equation}

By expanding the commutator and reordering terms,
the first order term can be rewritten
as
\begin{equation}
  T^{(1)}(\alpha)
    = - \tr\left( [\alpha a^\dagger - \alpha^* a, \Upsilon^\mathrm{target}] \rho \right) .
\end{equation}
This form is more convenient since it leads to commutators that can be precomputed. 
Indeed, so expressed, the first order term
can be further
explicited
as
\begin{equation}
  T^{(1)}(\alpha) = - \alpha \, \tr\left( [a^\dagger, \Upsilon^\mathrm{target}] \rho \right)
                    + \alpha^* \, \tr\left( [a, \Upsilon^\mathrm{target}] \rho \right) .
\end{equation}
Defining
\begin{equation}
  B = [a, \Upsilon^\mathrm{target}] \rho ,
\end{equation}
then
\begin{equation}
  B^\dagger = \rho^\dagger [a, \Upsilon^\mathrm{target}]^\dagger
            = - \rho [a^\dagger, \Upsilon^\mathrm{target}] ,
\end{equation}
%
hence
\begin{equation}
  \tr(B^\dagger) = - \tr(\rho [a^\dagger, \Upsilon^\mathrm{target}])
                   = - \tr( [a^\dagger, \Upsilon^\mathrm{target}] \rho ) .
\end{equation}
Now, since $\tr(B^\dagger) = \left( \tr B \right)^*$, then
\begin{equation}
  \tr{( [a^\dagger, \Upsilon^\mathrm{target}] \rho )} = - \left( \tr{(B)} \right)^* .
\end{equation}
With these developments, and setting
\begin{equation}
  \zeta = \tr B = \tr \left( [a, \Upsilon^\mathrm{target}] \rho \right) ,
\end{equation}
$T^{(1)}(\alpha)$ can simply be rewritten as
\begin{equation}
  T^{(1)}(\alpha) = \alpha \zeta^* + \alpha^* \zeta .
\end{equation}
It is convenient for the sequel to define the commutator
\begin{equation}
  C^{\Upsilon^\mathrm{target}} = [a, \Upsilon^\mathrm{target}] ,
\end{equation}
which can be precomputed;
with this
\begin{equation}
  B = C^{\Upsilon^\mathrm{target}} \rho
\end{equation}
and
\begin{equation}
  \zeta = \tr (C^{\Upsilon^\mathrm{target}} \rho) .
\end{equation}

Now, by similar reasoning as for $T^{(1)}(\alpha)$,
$T^{(2)}(\alpha)$
can be rewritten as
\begin{align}
  T^{(2)}(\alpha)
    &= \tr\left(
            \left[ \alpha a^\dagger - \alpha^* a,
                   [\alpha a^\dagger - \alpha^* a, \Upsilon^\mathrm{target}]
            \right] \rho
          \right) \notag \\
    &= \tr(K \rho) ,
\end{align}
where
\begin{equation}
  K = \left[ \alpha a^\dagger - \alpha^* a,
             [\alpha a^\dagger - \alpha^* a, \Upsilon^\mathrm{target}]
      \right] .
\end{equation}
%
Developing this operator, and using Jacobi's identity
$  \left[ A , [ B , C ] \right]
 + \left[ C , [ A , B ] \right]
 + \left[ B , [ C , A ] \right] = 0$
along with $[ a , a^\dagger] = \mathbb{I}$, one obtains
\begin{align}
  K &=  \alpha^2 \left[ a^\dagger , [ a^\dagger , \Upsilon^\mathrm{target}] \right]
     + {\alpha^*}^2 \left[ a , [ a , \Upsilon^\mathrm{target}] \right] \notag \\
    &\mathrel{\phantom{=}}
     -  2 \abs{\alpha}^2 \left[ a^\dagger , [ a , \Upsilon^\mathrm{target}] \right] .
\end{align}
Setting
\begin{equation}
  G^{\Upsilon^\mathrm{target}}
    = \left[ a , [ a , \Upsilon^\mathrm{target}] \right]
    = [ a , C^{\Upsilon^\mathrm{target}} ] ,
\label{eq:DefnGUpsilonTag}
\end{equation}
\begin{equation}
  E^{\Upsilon^\mathrm{target}}
    = \left[ a^\dagger , [ a , \Upsilon^\mathrm{target}] \right]
    = [ a^\dagger , C^{\Upsilon^\mathrm{target}} ] ,
\label{eq:DefnEUpsilonTag}
\end{equation}
and using that
\begin{equation}
  \left( G^{\Upsilon^\mathrm{target}} \right)^\dagger
    = \left[ a^\dagger , [ a^\dagger , \Upsilon^\mathrm{target}] \right] 
\end{equation}
leads to
\begin{align}
  K = \alpha^2 \left( G^{\Upsilon^\mathrm{target}} \right)^\dagger
     + {\alpha^*}^2 G^{\Upsilon^\mathrm{target}}
     -  2 \abs{\alpha}^2 E^{\Upsilon^\mathrm{target}} , 
\end{align}
and hence
\begin{align}
  T^{(2)}(\alpha)
    &= \tr( K \rho ) \notag \\
    &= \alpha^2 \tr\left( \left( G^{\Upsilon^\mathrm{target}} \right)^\dagger \rho \right)
     + {\alpha^*}^2 \tr\left( G^{\Upsilon^\mathrm{target}} \rho \right) \notag \\
    &\mathrel{\phantom{=}}
    -  2 \abs{\alpha}^2 \tr\left( E^{\Upsilon^\mathrm{target}} \rho \right) . \label{eq:T2alpha}
\end{align}
It
can
be shown that $T^{(2)}(\alpha)$ is a real quantity (the development will not be provided here).
%
\begin{align*}
  \tr\left( \left( G^{\Upsilon^\mathrm{target}} \right)^\dagger \rho \right)
  &= \tr\left( \left( G^{\Upsilon^\mathrm{target}} \right)^\dagger \rho^\dagger \right) \\
  &= \tr\left( \left( \rho G^{\Upsilon^\mathrm{target}} \right)^\dagger  \right) \\
  &= \tr\left( \left( \rho G^{\Upsilon^\mathrm{target}} \right) \right)^*
\end{align*}
Defining
\begin{equation}
  \gamma = \tr\left( G^{\Upsilon^\mathrm{target}} \rho \right) ,
\end{equation}
and
\begin{equation}
  \chi = \tr\left( E^{\Upsilon^\mathrm{target}} \rho \right) ,
\end{equation}
%
where $\chi$ can be demonstrated to be real,
$T^{(2)}(\alpha)$ can be written as
\begin{equation}
  T^{(2)}(\alpha) = \alpha^2 \gamma^* + {\alpha^*}^2 \gamma
                      - 2 \abs{\alpha}^2 \chi . 
\end{equation}
%

With the previous developments, the second order expansion of the Lyapunov function can be rewritten as (refer back to Eq.~\eqref{eq:LyapunovFct2ndOrder})
\begin{align}
  V_\Upsilon^\mathrm{target}(\rho')
    &\approx
      V_\Upsilon^\mathrm{target}(\rho)
      + \alpha \zeta^* + \alpha^* \zeta \notag \\
    &\mathrel{\phantom{\approx}} \mathrel{\phantom{=}}
      + \frac{1}{2}
        \left(
          \alpha^2 \gamma^* + {\alpha^*}^2 \gamma - 2 \abs{\alpha}^2 \chi
        \right)  \notag \\
    &\mathrel{\phantom{=}} =
      V_\Upsilon^\mathrm{target}(\rho) + q(\alpha) ,
\label{eq:LyapunovFct2ndOrderQuadForm}
\end{align}
with $q(\alpha)$ being the following quadratic form:
\begin{equation}
  q(\alpha)
    = \alpha \zeta^* + \alpha^* \zeta
      + \frac{1}{2}
          \left(
            \alpha^2 \gamma^* + {\alpha^*}^2 \gamma - 2 \abs{\alpha}^2 \chi
          \right) .
\label{eq:QuadFormLyapunovFct}
\end{equation}

Recall that it is required to determine $\alpha$ according to the inequality given in Eq.~\eqref{eq:IneqChoiceAlpha}, which means that $q(\alpha)$ must be negative, \textit{i.e.}
\begin{equation}
  q(\alpha) < 0,
\label{eq:NegQuadForm}
\end{equation}
and ideally $q(\alpha)$ shall be as negatively large as possible.

There are different ways in which $q(\alpha) < 0$. One standard possibility is equivalent to gradient steepest descent, and a second one is equivalent to a Newton method in numerical optimization, which is resorted to,
as it has faster convergence properties.

\subsection{Newton method}

The
quadratic form $q(\alpha)$
is
first be represented in terms of real quantities. First, $q(\alpha)$ is written as follows:
\begin{equation}
  q(\alpha)
    = 2 \Re{\alpha \zeta^*}
      + \Re{\alpha^2 \gamma^*} - \abs{\alpha}^2 \chi .
\end{equation}
The complex quantities appearing in $q(\alpha)$ are written as
\begin{align}
  \alpha &= x + i y, \\
  \zeta &= u + i v, \\
  \gamma &= g + i h,
\end{align}
This allows writing the quadratic form as
\begin{align}
  q(\alpha) \equiv q(x, y)
  &=
    \left[
      \begin{array}{cc}
        x & y
      \end{array}
    \right]
    Q
    \left[
      \begin{array}{c}
        x \\
        y
      \end{array}
    \right]
    +
    2 L
    \left[
      \begin{array}{c}
        x \\
        y
      \end{array}
    \right] ,
\end{align}
with
\begin{equation}
  Q = \left[
        \begin{array}{cc}
          g - \chi & h \\
             h     & -(g + \chi)
      \end{array}
    \right]
  , \quad
  L = \left[
        \begin{array}{cc}
          u & v
        \end{array}
      \right] .
\end{equation}
The Newton approach in optimization is to take the direction
$\left[ \begin{array}{cc} x & y \end{array} \right]^\mathrm{T}$ which minimizes the quadratic form. This leads to
\begin{equation*}
  Q
  \left[
    \begin{array}{c}
      x \\
      y
    \end{array}
  \right]
  = - L^\mathrm{T} ,
\end{equation*}
which is equivalent to
\begin{equation}
  \left[
    \begin{array}{c}
      x \\
      y
    \end{array}
  \right]
  = - Q^{-1} L^\mathrm{T} .
\label{eq:NewtonDirxy}
\end{equation}
The inverse of $Q$ is easily obtained and given by
\begin{equation}
  Q^{-1}
  =
  \frac{1}{(g^2 + h^2 - \chi^2)}
  \left[
    \begin{array}{cc}
      g + \chi & h \\
      h     & -(g - \chi)
    \end{array}
  \right] .
\end{equation}
Re-expressing everything in term of the complex quantities $\zeta$, $\gamma$, and $\chi$ gives
\begin{equation}
  \left[
    \begin{array}{c}
      x \\
      y
    \end{array}
  \right]
  = - Q^{-1} L^\mathrm{T}
  =
  \frac{1}{\chi^2 - \abs{\gamma}^2}
  \left[
    \begin{array}{c}
      \Re{\chi \zeta + \gamma \zeta^*} \\
      \Im{\chi \zeta + \gamma \zeta^*} 
    \end{array}
  \right] ,
\end{equation}
hence
\begin{equation}
  \alpha
    = \frac{1}{\chi^2 - \abs{\gamma}^2}
      (\chi \zeta + \gamma \zeta^*) .
\label{alpha_newton}
\end{equation}

Equation~\ref{alpha_newton} is therefore the one defining the update rule for the control amplitude parameter in the feedback procedure. For Newton's method, it is necessary that the matrix $Q$ be positive-definite~\cite{Fletcher_2000}. This is true if both eigenvalues of $Q$ are positive. These eigenvalues are easily found to be
\begin{equation}
  \lambda_\pm = - \chi \pm \abs{\gamma} .
\end{equation}
Hence, for $Q$ to be positive-definite the following must hold:
\begin{equation}
  \chi < -\abs{\gamma} .
\end{equation}
This condition is numerically verified.
For the values of $x$ and $y$ given through Eq.~\eqref{eq:NewtonDirxy}, the value of $q(x, y)$ is found to be
\begin{equation}
  q(x, y) = - L \, Q^{-1} L^\mathrm{T} ,
\end{equation}
which is negative whenever $Q$ is positive-definite, since in this case $Q^{-1}$ is also positive definite.

\bibliography{apssamp_zotero}
    
\end{document}